\documentclass[twocolumn,showpacs,showkeys,amsmath,amssymb,superscriptaddress,floatfix,nofootinbib]{revtex4}

\usepackage{graphicx}
\usepackage{bm} 
\usepackage{subfigure}
	
\def\vec#1{\mathchoice{\mbox{\boldmath$\displaystyle#1$}}
{\mbox{\boldmath$\textstyle#1$}}
{\mbox{\boldmath$\scriptstyle#1$}}
{\mbox{\boldmath$\scriptscriptstyle#1$}}}
\makeatletter

\begin{document}
 
\title{Non-identical particle femtoscopy at $\sqrt{s_{NN}}=200$~GeV in
hydrodynamics with statistical hadronization%
\footnote{Supported by the U.S. NSF Grant No. PHY-0653432.}}

\author{Adam Kisiel} 
\email{kisiel@if.pw.edu.pl}
\affiliation{Department of Physics, Ohio State University, 
1040 Physics Research Building, 191 West Woodruff Ave., 
Columbus, OH 43210, USA}
\affiliation{Faculty of Physics, Warsaw University of Technology, PL-00661 Warsaw, Poland}

 
\begin{abstract}
Non-identical particle femtoscopy probes not only the size of the
emitting system, but also the emission asymmetries between particles
of different mass, which are intimately related with the collective
behavior of matter. We apply the technique to the simulations from the
{\tt THERMINATOR}+Lhyquid model of the heavy-ion collisions at
$\sqrt{s_{NN}}=200$~GeV. We present predictions for all pairwise
combinations of pions, kaons, and protons, and discuss their
interpretation. We show that kaon and proton distributions are
strongly influenced by flow: the source gets smaller and shifted to
the outside with growing $p_T$,  while for pions the shift is
significantly smaller, producing an emission asymmetry. We explain how
particles coming from decays of hadronic resonances enhance the
asymmetry signal coming from flow, contrary to naive
expectations. Emphasis is put on extracting this unique information
about collective behavior of matter from the non-identical particle
correlations. We also present, in detail, the  technical aspects of
the non-identical particle femtoscopy technique applied to data from
the heavy-ion collisions. We list the sources of systematic errors
coming from the method itself and the usual assumptions. We describe
robust analysis methods and discuss their limitations.   

\end{abstract}

\pacs{25.75.-q, 25.75.Dw, 25.75.Ld}

\keywords{relativistic heavy-ion collisions, hydrodynamics,
femtoscopy, non-identical particle correlations, RHIC, LHC}

\maketitle 


\section{Introduction}
\label{sec:intro}

Femtoscopy has been used for more than 35
years~\cite{Kopylov:1972qw,Kopylov:1973qq} to measure sizes of
the systems created in nucleus-nucleus collisions. Initially it was
developed for analyzing the two-particle correlation arising from the 
wave-function symmetrization for pairs of identical
particles~\cite{Goldhaber:1960sf} and was similar in mathematical
framework to the ``HBT interferometry'' used in
astronomy~\cite{HanburyBrown:1954wr,Brown:1956zz,HanburyBrown:1956pf}.
Later it was realized that similar correlations arise due to the Final
State Interactions (Coulomb and Strong) between particles that are not
necessarily
identical~\cite{Lednicky:1995vk,Voloshin:1997jh,Lednicky:2003fe,Lednicky:2005tb,Pratt:2006jf}.  

Femtoscopy of non-identical particles provides unique information:
while the identical particle correlation usually only measure 
the ``size'' of the emitting region (more precisely, the second
moments of the emission function), non-identical correlations can,
thanks to the very fact that they correlate particles that are not
identical, also measure the relative emission shifts (the first moments
of the emission
function)~\cite{Lednicky:1981su,Boal:1990yh,Lednicky:1995vk,Voloshin:1997jh,Lednicky:2003fe,Lednicky:2005tb,0954-3899-35-12-125109}.

The unique features of the non-identical particle femtoscopy was
used in low-energy nuclear collisions to study the time ordering of
the emission of various nuclear fragments from the compound
nucleus~\cite{Erazmus:1994xxx,Kotte:1999gr,Gourio:2000tn,Ghetti:2001gm,Helgesson:2005fk}.
This required measuring time differences from several to hundreds of
fm/c~\cite{Kotte:1999gr,Gourio:2000tn,Ghetti:2001gm,Helgesson:2005fk}.

In this work we focus on the applications of this technique in the
collisions of ultra-relativistic heavy ions, specifically Au ions at
the Relativistic Heavy Ion Collider (RHIC) in Brookhaven National
Laboratory (BNL), although the discussion is also relevant for
lower (at Super Proton Synchrotron SPS of CERN) and higher (Large
Hadron Collider LHC at CERN) energy collisions. 

One of the major discoveries at RHIC has been the observation of the
Quark Gluon Plasma (QGP), which was found to behave collectively, very
much like a 
fluid~\cite{Adams:2005dq,Adcox:2004mh}. Hydrodynamic equations seemed 
to describe this behavior well in the momentum sector. However
attempts to simultaneously describe the space--time behavior measured
by femtoscopy have not been successful until recently. This failure
was commonly referred to as the ``RHIC hydro HBT
puzzle''~\cite{Heinz:2002un,Hirano:2004ta,Lisa:2005dd,Huovinen:2006jp}.
Through a 
detailed analysis of the experimental data and various improvements of
the hydrodynamic description it was realized that certain
assumptions had to be modified in order to properly describe both
sets of observables. The initial condition was changed to the one that
uses Gaussian initial profile for the transverse energy
density~\cite{Broniowski:2008vp}, which results in faster development
of the initial flow, compared the traditional
one~\cite{Sinyukov:2006dw,Sinyukov:2009xe}. The  
equation-of-state used did not exhibit a first-order phase transition,
but rather a cross-over. The detailed simulation of the resonances
contribution in the later stage of the 
collisions was carried out~\cite{Broniowski:2008vp}. In addition, 
some studies suggest that the introduction of viscosity and universal
(pre-equilibrium flow) into the
model may also play an important role~\cite{Pratt:2008bc,Pratt:2008sz}. 

The hydrodynamic scenario produces specific space-momentum correlation
patterns, which are commonly referred to as flow. The system created
in the heavy-ion collision expands rapidly outwards, showing a very
strong radial flow, which is observed in the modification of the
single particle inclusive $p_{T}$ spectra shape. In addition,
in non-central collisions, the initial overlap region has an elongated
(usually described as ``almond'') shape, and this spatial
asymmetry is converted to the momentum one in the final state. This
momentum asymmetry is observed as an elliptic flow $v_{2}$ and is the
subject of very intensive theoretical and experimental studies;
see~\cite{Voloshin:2008dg,Sorensen:2009cz} for recent reviews. However,
because of its origin, it is small in central collisions. This is
unfortunate, since it is in central collisions that we expect to
create the largest volume of deconfined matter, which we would
like to study. Both the $p_{T}$ spectra and the elliptic flow are
observables depending only on the momenta of the particles,
so their connection to space--time 
can only be indirectly inferred. To access it directly we employ
femtoscopic techniques. It is argued that the fall of the ``femtoscopic radii''
with particle's $m_T$ can be interpreted as the decrease of ``lengths
of homogeneity'', a direct consequence of radial and longitudinal
flow~\cite{Akkelin:1995gh}. However, one might come up with alternative
explanations, involving temperature gradients to produce similar
dependencies~\cite{Gombeaud:2009fk}. In this work we will show how the
collective flow present in hydrodynamic and transport models, in
addition to the effects discussed above, produces differences in
average emission points between particles of different masses. We will
also describe
how these emission asymmetries can be accessed via the non-identical 
particle femtoscopy~\cite{Lednicky:1999nn}. Measuring such effects
would enable us to eliminate 
scenarios alternative to hydrodynamic expansion and provide a crucial
and strict test for the models. Results on $\pi^{+}$--$\pi^{-}$ and
$\pi$--proton correlations have been previously reported by
NA49~\cite{Lednicky:1999nn,Blume:2002mr} and 
the CERES experiment~\cite{Antonczyk:2007id,Miskowiec:1998ms} at the
SPS. The first results on pion--kaon correlations have been reported
by STAR~\cite{Adams:2003qa}. Preliminary
results on $\pi$--$\Xi$ correlations from STAR have also been
presented~\cite{Chaloupka:2007gw}. Proton--$\Lambda$ correlations have
also been measured by STAR~\cite{Adams:2005ws}, but this analysis did
not attempt to extract emission asymmetries. 

In order to properly simulate the emission asymmetries between
particles of different masses we need a model which has all the
important features: the hydrodynamic phase which produces
space-momentum correlations and the hadronic phase where at least the
hadronic resonance decays and propagation is treated. In addition the
model should provide the space--time freeze-out coordinates of
particles, so that femtoscopic calculations can be carried out. It
should, as much as possible, reproduce the available data on particle
spectra and femtoscopy. Hydrodynamics inspired ``blast-wave''
parametrization of 
freeze-out have been used to model asymmetries between non-identical
particles~\cite{Retiere:2003kf}. This study nicely illustrated the
connection between strong collective matter behavior and the
asymmetries; however it used simplified emission functions and
neglected the resonance propagation and decay. In this 
work we have chosen the {\tt THERMINATOR}+Lhyquid model, which we
introduce in Section~\ref{sec:therminator}. In 
Section~\ref{sec:modelpreditions} we describe the particular set of
simulations of heavy-ion collisions at six centralities that we have
carried out. We calculated the system size and emission asymmetry for
three pair types: pion--kaon, pion--proton and kaon--proton. We discuss
the origins of the asymmetry and provide the connection to the 
space-momentum correlations coming from flow. We also discuss other
non-flow sources of emission asymmetries and provide a quantitative
estimate of all contributions. 

Having emphasized the importance of the emission asymmetries
phenomenon we proceed to describe the theoretical framework of the
non-identical particle correlations in
Section~\ref{sec:nonidform}. Then, we move to the technical aspects of 
the measurement in heavy-ion collisions in
Section~\ref{sec:implement}. We argue that a specific mathematical 
representation of the non-identical correlation functions (the spherical
harmonics decomposition) shows remarkable synergies with the analysis
and  maximizes the statistical significance of the emission asymmetry 
signal. We describe the measurement procedure which enables to recover
the properties of the model emission function from the
``experimental'' correlation function. In
Section~\ref{sec:crosschecks} we test the robustness of the
procedure, list the assumptions and the approximations that need to be
employed and estimate the systematic error coming from the technique
itself. We show that the most important contribution to this error is
the correct estimation of the fraction of non femtoscopically
correlated pairs (traditionally called ``purity''). We analyze this
effect in detail and present purity estimates, based on our model of
choice, which can be directly used in the experimental analysis.

Finally, in Section~\ref{sec:experimental} we use numerical simulations
with the {\tt THERMINATOR} model to produce ``experimental-like''
correlation functions, which we then analyze with the methods
presented earlier. We demonstrate that the method is able to recover
the theoretical input values. The results presented in this
section are the theoretical predictions for the overall system size as
well as the emission asymmetry for all considered pair types and 6
centralities. They can be immediately compared to the experimental
results with minimal number of approximations, and therefore minimal
systematic uncertainty.

\section{Lhyquid + {\tt THERMINATOR} model}
\label{sec:therminator}

In this work we use what we call the {\em standard approach},
consisting of ideal-fluid hydrodynamics followed by statistical
hadronization. Numerous calculations have been performed in this
framework, with the  common difficulty~\cite{Heinz:2002un} of
simultaneously describing femtoscopy and other signatures in the
data. More precisely, the {\em RHIC HBT
puzzle}~\cite{Heinz:2002un,Hirano:2004ta,Lisa:2005dd,Huovinen:2006jp}
refers 
to problems in reconciling the large value of the elliptic flow
coefficient, $v_2$, with the identical particle interferometry in
calculations based on hydrodynamics 
~\cite{Heinz:2001xi,Hirano:2001yi,Hirano:2002hv,Zschiesche:2001dx,Socolowski:2004hw}.
Recently, a successful uniform description of soft observables at RHIC
has been accomplished, including the femtoscopic radii, within the standard
approach~\cite{Broniowski:2008vp}. The essential ingredients of this
analysis are the Gaussian initial condition for hydrodynamics, early
start of the evolution, the state-of-the art equation of state with
smooth crossover, and the use of {\tt THERMINATOR}~\cite{Kisiel:2005hn} with all resonances from {\tt SHARE}~\cite{Torrieri:2004zz} incorporated to carry out the statistical
hadronization at the freeze-out surface of temperature $150$~MeV. The
interplay of these elements resulted in a simultaneous description of
the transverse-momentum spectra of pions, kaon and protons, the $v_2$,
and the femtoscopic correlation radii of pions, including full
centrality, $k_T$ and reaction plane dependence of azimuthally
sensitive HBT signatures~\cite{Kisiel:2008ws}.  

In this section we describe the essential elements of our method to the extent they are necessary 
for the presented new results.  
More details concerning the hydrodynamics can be found in 
Refs.~\cite{Chojnacki:2006tv,Chojnacki:2007jc,Broniowski:2008vp}, while 
the method used for femtoscopic calculations has been presented in Ref.~\cite{Kisiel:2006is}.

\subsection{Initial condition}

As reported in Ref.~\cite{Broniowski:2008vp}, the use of the initial
condition for hydrodynamics of the Gaussian form,
\begin{eqnarray}
n(x,y)=\exp \left ( -\frac{x^2}{2a^2} -\frac{y^2}{2 b^2} \right ),
\label{profile}
\end{eqnarray} 
where $n$ is the initial energy density, while $x$ and $y$ denote the
transverse coordinates, leads to a much 
better uniform description of the data for the $p_T$-spectra, $v_2$,
and the pionic femtoscopic radii compared to the use of the standard initial
condition from the Glauber model.  

The width parameters $a$ and $b$ depend on centrality. In order to
estimate realistic values for them we run the {\tt
GLISSANDO}~\cite{Broniowski:2007nz} Glauber Monte Carlo simulations
which include the eccentricity 
fluctuations~\cite{Andrade:2006yh,Hama:2007dq}. Then we match $a^2$
and $b^2$ to reproduce the values $\langle x^2 \rangle$ and $\langle
y^2 \rangle$ from the {\tt GLISSANDO} profiles. Thus, by construction,
the spatial RMS radii of the initial condition and its eccentricity is
the same as the ones from the Glauber calculation. Nevertheless, the
shape is not the same. The Gaussian profiles are sharper near the
origin, which results in a faster buildup of the Hubble-like flow in
the hydrodynamical stage.

The Glauber calculations, needed to obtain the $a$ and $b$ parameters,
correspond to the mixed model~\cite{Kharzeev:2000ph}, where the number
of produced particles is proportional to $(1-\alpha)N_w/2 + \alpha
N_{\rm bin}$, with $N_w$ and $N_{\rm bin}$ denoting the number of
wounded nucleons~\cite{Bialas:1976ed} and binary collisions,
respectively. The parameter $\alpha=0.145$ for top RHIC energy
~\cite{Back:2001xy,Back:2004dy}. The inelastic nucleon cross section is
42~mb for RHIC~\cite{Yao:2006px}. The simulations incorporate the
fluctuations of orientation of the fireball (the variable-axes
geometry), which result in increased eccentricity compared to the
fixed-axes geometry~\cite{Broniowski:2007ft}. Finally, the expulsion
distance of 0.4~fm is used in the generation of the nuclear
distributions. A source-dispersion parameter of 0.7~fm is
used. It describes the random displacement of the source
from the center of the wounded nucleon or the binary-collision
position~\cite{Broniowski:2007nz}.  

The values of the $a$ and $b$ parameters for various centralities and
the corresponding eccentricity parameters 
\begin{eqnarray}
\epsilon^\ast=\frac{b^2-a^2}{a^2+b^2}, \label{eps}
\end{eqnarray}
are collected in Table~\ref{tab:ab}. 

\begin{table*}[tb]
\caption{Shape parameters $a$ and $b$ of Eq.~(\ref{profile}) for
various centrality classes obtained by matching  $\langle x^2 \rangle$
and $\langle y^2 \rangle$ to  {\tt GLISSANDO} simulations, the
variable-axes eccentricity $\epsilon^\ast$, and the central
temperature $T_i$. 
\label{tab:ab}}
\begin{tabular}{|r|rrrrrrrrr|}
\hline
$c$ [\%] & 0-5  & 5-10 & 10-20 & 20-30 & 30-40 & 40-50 & 50-60 & 60-70 & 70-80\\
\hline
 & \multicolumn{9}{|c|}{RHIC Au+Au at $\sqrt{s_{NN}}=200$~GeV}\\
$a$ [fm] & 2.70 & 2.54 & 2.38  & 2.00  & 1.77  & 1.58  & 1.40  & 1.22  & 1.04 \\
$b$ [fm] & 2.93 & 2.85 & 2.74  & 2.59  & 2.45  & 2.31  & 2.16  & 2.02  & 1.85 \\
$\epsilon^\ast$ & 0.08 & 0.12 & 0.18 & 0.25 & 0.31  & 0.36  & 0.41  & 0.46  & 0.52 \\
$T_i$ [MeV] & 500  & 491     & 476      & 460      & 429       & 390      & 344      & 303      & 261     \\
\hline
\end{tabular}
\end{table*}

The energy-density profile (\ref{profile}) determines the initial
temperature profile via the equation of
state~\cite{Chojnacki:2007jc}. The initial central temperature, $T_i$,
is a parameter dependent on centrality. For RHIC calculations it is 
adjusted to reproduce the total particle multiplicity.   

\subsection{Hydrodynamics}

The hydrodynamics equations used in this work were described in
greater detail in Refs.~\cite{Chojnacki:2006tv,Chojnacki:2007jc}. We
use inviscid (ideal-fluid), baryon-free, boost-invariant
hydrodynamics. The equations are written in terms of the velocity of
sound, $c_s$, whose temperature dependence encodes the full
information on the equation of state of the system. We
incorporate the known features of $c_s(T)$, which, at high temperatures,
are given by the lattice QCD calculations~\cite{Aoki:2005vt}; at low $T$ they follow from the
hadron gas including all resonances, while at intermediate $T$ an
interpolation is used. No sharp phase transition, but a smooth
cross-over, is built in, in accordance to the present knowledge of the
thermodynamics of QCD at zero baryon chemical potential. The plot of
the resulting $c_s(T)$ can be found in  Ref.~\cite{Broniowski:2008vp}. 

The initial proper time of the start of hydrodynamics is fixed to
the value 
\begin{eqnarray}
\tau_0=0.25{\rm ~fm}.
\end{eqnarray}
This early start of hydrodynamics allows for a fast generation of
transverse flow, similar to the effect described in~\cite{Pratt:2008sz}. 

\subsection{Freeze-out}

The hydrodynamic evolution proceeds until freeze-out occurs, where the
assumed condition for the {\em universal} freeze-out temperature is
$T_f=150$~MeV. This value is somewhat lower than in several fits of
the chemical
freeze-out~\cite{BraunMunzinger:2001ip,Broniowski:2001we,Broniowski:2001uk};
however, it agrees with the recently made global fits to particle
transverse momentum spectra of
Ref.~\cite{Rafelski:2004dp,Prorok:2007xp}, where the value around
150~MeV was obtained for the kinetic freeze-out
temperature.\footnote{The use of this lower freeze-out temperature
needs the introduction of the strangeness inequilibrium factors
$\gamma_s$ in order to reproduce the abundances of  strange
particles~\cite{Rafelski:2004dp}.} 


{\tt THERMINATOR} is used to carry out the statistical hadronization
at the freeze-out hypersurface according to the Cooper-Frye
formulation~\cite{Cooper:1974mv}. According to the assumed
single-freeze-out approximation, identifying the kinetic and chemical
freeze-out temperatures, rescattering processes after freeze-out are
neglected. We have checked that the collision rate after freeze-out is
moderate for the hypersurfaces applied in this work. We estimate it by
considering a pion straight-line trajectory and counting the number of
encounters with other particles closer than the distance corresponding
to the pion-hadron cross section. The average number of these
trajectory crossings is about 1.5-1.7 per pion. This shows that the
single-freeze-out approximation~\cite{Broniowski:2001we} works
reasonably well for the present case. At a more detailed level, one
could use hadronic afterburners to model the elastic
collisions~\cite{Teaney:2000cw,Nonaka:2006yn,Hirano:2007xd}, or
attempt the 
hydro-kinetic approach implemented
in~\cite{Sinyukov:2006dw,Sinyukov:2009xe,Amelin:2006qe}.

\subsection{Two-particle femtoscopy}

The method used for femtoscopic analysis of the {\tt THERMINATOR}
model output was described in detail in Ref.~\cite{Kisiel:2006is}. The
features of the 
formalism specific to the non-identical particle correlations are
described in Section~\ref{sec:nonidform}. Their discussion is one of
the main points of this work.  

\section{Model predictions for RHIC}
\label{sec:modelpreditions}

In this section we discuss the general trends that are expected to
emerge in the non-identical correlations. We also discuss their
physical origins and the possible physics conclusions that can be
drawn when they are observed experimentally. Later in this paper, we
will see how these trends manifest themselves in the calculations and
we test whether it is indeed possible to observe them in the
experiment. 

\begin{figure}[tb]
\begin{center}
\includegraphics[angle=0,width=0.4 \textwidth]{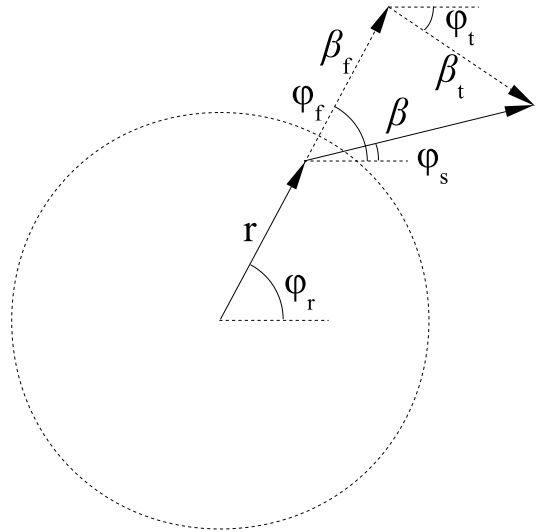}
\end{center}
\vspace{-6.5mm}
\caption{Diagram of the particle's velocity $\beta$ decomposition into
the flow $\beta_f$ and thermal $\beta_t$ components.
\label{fig:betaflow}}
\end{figure}

Following convention, we assign symbol $x$ for the emission points of
single particles, usually used as a vector, $\vec x$, or as a vector
magnitude, $x$. Momentum of a particle is denoted as $p$. Single
particle velocity is $\beta$. We also define the pair
variables. Relative separation between particles (vector) is: $\vec r
= \vec x_1 - \vec x_2$. We use the out-side-long coordinate system,
where the $long$ or $longitudinal$ direction is along the beam axis,
the $out$ or $outward$ direction is along the pair total transverse
momentum and the $side$ or $sideward$ is perpendicular to the other
two. In Longitudinally Co-Moving System (LCMS) the pair longitudinal
momentum vanishes: $p_{1,long} = -p_{2,long}$. In the Pair Rest Frame
(PRF) (also called Pair Center of Mass (PCOM)) the pair center-of-mass
rests: $\vec p_1 = -\vec p_2$. In our convention all pair variables in
PRF are marked by an asterisk. The pair relative momentum half is
denoted as $k^{*}$ and is equal to the {\it first} particle's momentum
in PRF. The pair total momentum is denoted as $P$ (or sometimes $K$
traditionally used in identical particle femtoscopy). All
single-particle coordinates as well as pair relative variables use
lower case letters. In contrast we will use upper-case letters to
denote the parameters of distributions. In particular we call the
variance or the two-particle separation distribution $R$. We call the
mean of such distribution $\mu$ (with the $L$ superscript if they are
defined in LCMS). We refer the reader to Appendix~\ref{sec:emfuncs}
for explicit mathematical formulas as well as relations between source
characteristics in the two reference frames. 

The discussion below is based on an example calculation from the {\tt
THERMINATOR}+Lhyquid model, done for parameters tuned to the central
(0-5\%) RHIC Au+Au collisions at $\sqrt{s_{NN}} = 200$~GeV. Whenever
we mention RHIC Au+Au calculations we mean simulations at the top RHIC
energy.

\subsection{Emission asymmetries}

\begin{table*}[tb]
\caption{Single-particle source parameters (mean $\left < r
\right >$ and RMS $X$) for three
particle types in selected kinematic regions.   
\label{tab:sizes}}
\begin{tabular}{|r|c|c|cc|cc|}
\hline
Particle & $\beta$ & $p_{T}$ [GeV/c] & $X_{out}$ [fm] & $\left <
x_{out} \right >$ [fm] &
$X_{side}$ [fm] & $\left < x_{side} \right >$ [fm] \\
\hline
 \multicolumn{7}{|c|}{primordial}\\
\hline
K     & 0.6-0.8   & 0.42-0.56 & 2.68 & 4.46  & 3.25 &  0.03 \\
$\pi$ & 0.6-0.8   & 0.12-0.16 & 3.47 & 2.85  & 3.66 & -0.02 \\
$\pi$ & 0.95-0.97 & 0.42-0.56 & 2.56 & 4.81  & 3.31 & -0.01 \\
\hline
p     & 0.6-0.8     & 0.80-1.06 & 2.02 & 5.40 & 2.82 & -0.05 \\
$\pi$ & 0.6-0.8     & 0.12-0.16 & 3.47 & 2.85 & 3.66 & -0.02 \\
$\pi$ & 0.985-0.991 & 0.80-1.06 & 1.88 & 5.68 & 2.90 &  0.00 \\
\hline
 \multicolumn{7}{|c|}{all}\\
\hline
K     & 0.6-0.8   & 0.42-0.56 & 2.99 & 4.97 & 3.55 &  0.00 \\
$\pi$ & 0.6-0.8   & 0.12-0.16 & 4.52 & 1.73 & 4.72 & -0.03 \\
$\pi$ & 0.95-0.97 & 0.42-0.56 & 2.94 & 5.03 & 3.20 & -0.01 \\
\hline
p     & 0.6-0.8     & 0.80-1.06 & 2.31 & 6.12 & 2.88 & -0.02 \\
$\pi$ & 0.6-0.8     & 0.12-0.16 & 4.52 & 1.73 & 4.52 & -0.02 \\
$\pi$ & 0.985-0.991 & 0.80-1.06 & 2.19 & 6.20 & 3.08 & -0.01 \\
\hline
\end{tabular}
\end{table*}

\begin{table*}[tb]
\caption{Mean emission points and pair asymmetries for pions, kaons
and protons with velocity $(0.6,0.8)$ in central Au+Au
simulation. Averaging is done over all pairs in this range. See
text for explanation of ``flow'' and ``scrambled''.
\label{tab:shifts}}
\begin{tabular}{|r|c|c|c|c|c|c|}
\hline
& $ \left < x_{out}^{\pi} \right > $ & $ \left < x_{out}^{K} \right > $ &
$ \left < x_{out}^{p} \right > $ & $ \left < x_{out}^{\pi} \right > -
\left < x_{out}^{K} \right > $ & 
$ \left < x_{out}^{\pi} \right > - \left < x_{out}^{p} \right >$ & $
\left < x_{out}^{K} \right > - \left < x_{out}^{p} \right > $
\\
\hline
 \multicolumn{7}{|c|}{flow}\\
\hline
primordial     & 2.83 & 4.47 & 5.61 & -1.64 & -2.78 & -1.14 \\
non-primordial & 1.34 & 7.35 & 9.19 & & & \\
all            & 2.00 & 5.54 & 6.69 & -3.54 & -4.69 & -1.15 \\
\hline
 \multicolumn{7}{|c|}{scrambled}\\
\hline
primordial     & -0.04 & 0.00 & -0.03 & -0.04 & -0.01 & 0.03 \\
non-primordial &  0.88 & 3.17 &  4.20 & & &\\
all            &  0.48 & 1.20 &  1.28 & -0.72 & -0.80 & -0.08 \\
\hline
\end{tabular}
\end{table*}

\begin{figure}[tb]
\begin{center}
\includegraphics[angle=0,width=0.48 \textwidth]{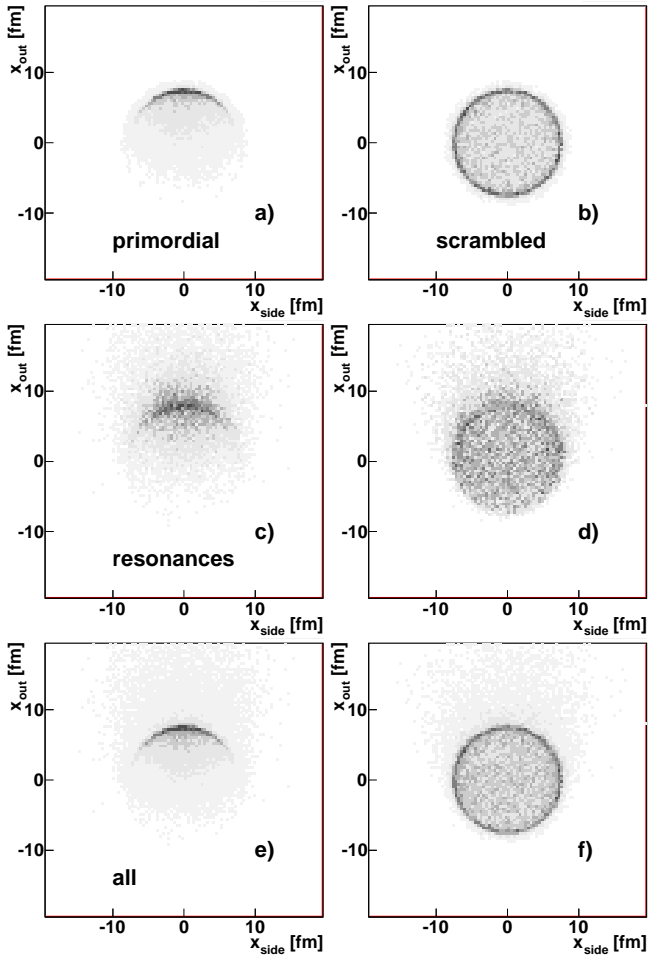}
\end{center}
\vspace{-6.5mm}
\caption{Emission points of particles with the velocity in
$(0.6,0.8)$, its direction pointing ``upwards'', for kaons, central
Au+Au. Panels a), c), e) show standard simulation results, panels b),
d), f): with flow correlation ``scrambled'' (see text for details). Upper
plots a), b) show primordial particles, center c), d): particles
coming from resonance decays, lower e), f): all. 
\label{fig:cankn}}
\end{figure}

\begin{figure}[tb]
\begin{center}
\includegraphics[angle=0,width=0.48 \textwidth]{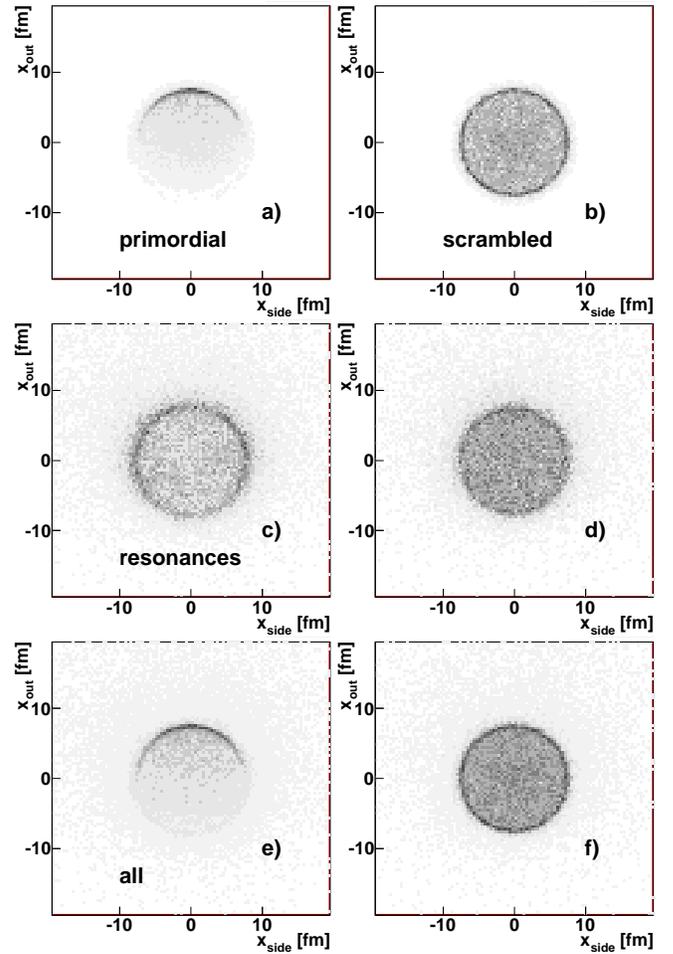}
\end{center}
\vspace{-6.5mm}
\caption{Emission points of particles with the velocity magnitude in
the range $(0.6,0.8)$, its direction pointing ``upwards'', for pions,
central Au+Au. Panels notation the same as in Fig.~\ref{fig:cankn}.
\label{fig:canpi}}
\end{figure}

\begin{figure}[tb]
\begin{center}
\includegraphics[angle=0,width=0.48 \textwidth]{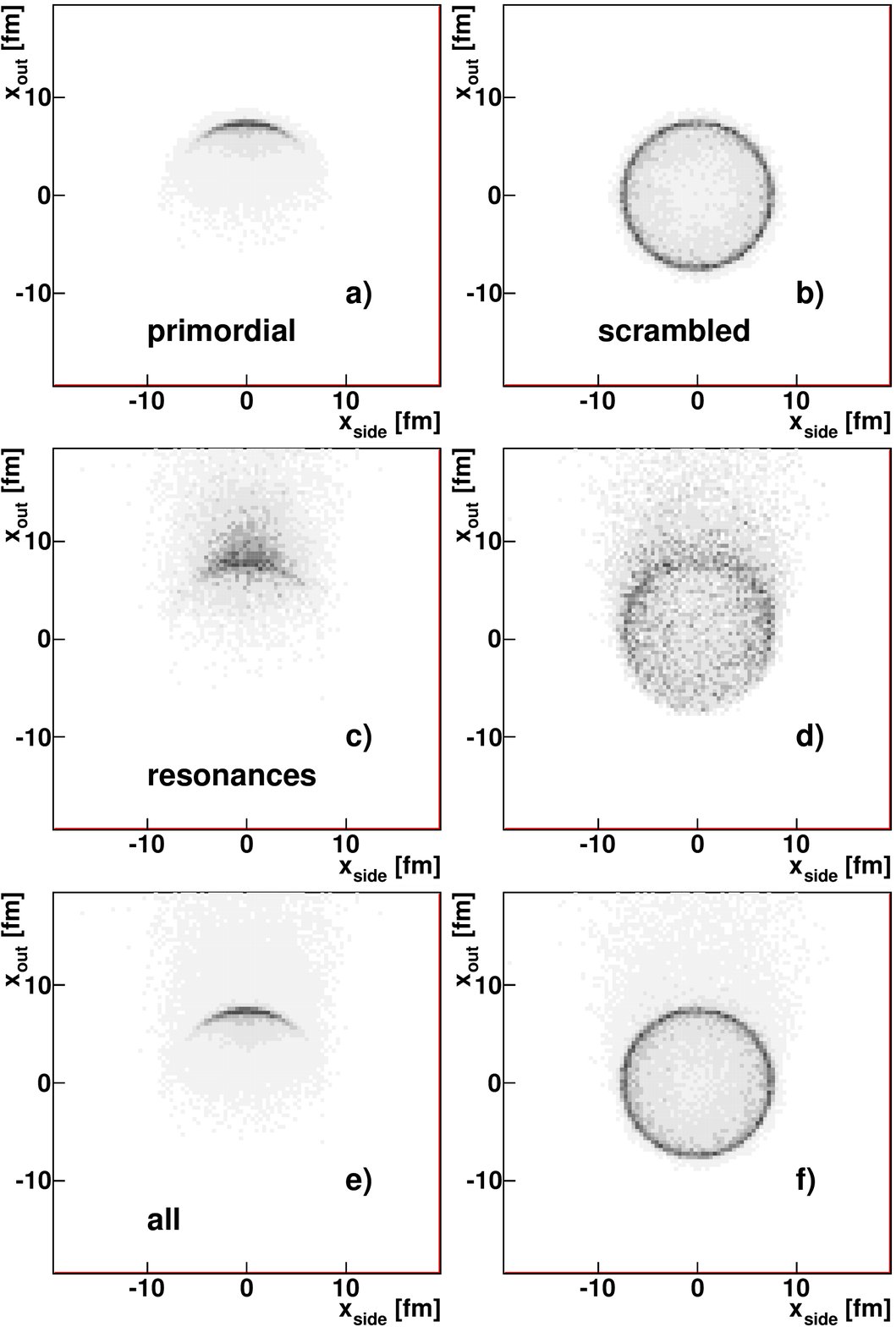}
\end{center}
\vspace{-6.5mm}
\caption{Emission points of particles with the velocity in
$(0.6,0.8)$, its direction pointing ``upwards'', for protons, central
Au+Au.  Panels notation the same as in Fig.~\ref{fig:cankn}. 
\label{fig:canpr}}
\end{figure}

Hydrodynamic evolution of matter implies strong space-momentum
correlations in particle emission. Particles emitted from a given
fluid cell will have a velocity which is a combination of two
components: the fluid cell velocity $\vec \beta_f$ (taken from the
flow field $u_{\mu}(\vec r)$) and the thermal velocity $\vec
\beta_t$. This is schematically shown in Fig.~\ref{fig:betaflow}.  

The system created in heavy-ion collision, when modeled in
hydrodynamics, naturally develops a collective behavior - the radial
flow in the transverse plane, that is matter is collectively moving
``outwards'' from the central axis of the source to the outside. In
essence this is an $x-p$ correlation - the direction $\phi_f$ of the
fluid element's transverse velocity is aligned with it's transverse
position vector direction $\phi_r$. When the fluid element emits
particles, all of them will have the same common flow velocity
$\beta_f$ taken directly from $u_{\mu}$. To this velocity one adds a
thermal component $\beta_t$ which has a random direction $\phi_t$ in
the rest frame of the fluid element. It will dilute the $x-p$
directional correlation. Let us now consider the mean emission point
of a single particle, more specifically its component parallel to the 
velocity~\cite{Lednicky:xderiv}: 
\begin{equation}
x_{out} = \frac {\vec x \vec \beta} {\beta} = \frac {r \left (
\beta_f + \beta_t \cos(\phi_t - \phi_f) \right )} {\beta} .
\label{eq:xoutfrombeta}
\end{equation}
We analyze its average over particles at fixed $\beta$. If we assume a
Gaussian density profile with radius $r_{0}$  and linear transverse
velocity profile $\beta_f = \beta_{0} r/r_{0}$ then we
obtain~\cite{Lednicky:xderiv}:   
\begin{equation}
\left < x_{out}  \right >  = \frac {\left < r \beta_f \right > }
{\left < \sqrt{\beta_t^2 + \beta_f^2} \right >} =
\frac {r_{0} \beta_0 \beta} { \beta_0^2 + T/m_{t} },
\label{eq:avxoutfrombeta}
\end{equation}
where we have explicitly given the formula for the velocity component
coming from temperature. Pions and kaons emitted from the fluid
element will have the same common flow velocity $\beta_f$ (and
different momenta). They will also get a random thermal kick,
depending on momentum. For a pion the same $p_{T}$ kick will mean much
larger {\it velocity} $\beta_t$ kick, than for a kaon. The final
velocity direction of a pion will be, on average, less correlated with
its emission position than that of a kaon. We assume that the
spatial characteristics of pion and kaon emission are the same and the
flow velocity is also the same. Hence, for both pions and kaons, $\left < r
\beta_f \right >$ is the same. The only difference is the $T/m_T$
component, smaller for kaons. Therefore, $\left < x_{out} \right
>$ for pions is smaller than $\left < x_{out} \right >$ for
kaons. Summarizing: when correlating pions and kaons with the same
velocity (or more generally two non-identical particles of different
masses, but same velocity) pions (lighter particles) appear to be, on
average, emitted closer to the center of the system than kaons
(heavier particles). Consequently, hydrodynamics predicts a negative
emission asymmetry $\mu_{out}^{light,heavy} = \left <
  r_{out}^{light,heavy} \right > = \left < x_{out}^{light} -
  x_{out}^{heavy} \right >$ between non-identical particles of
different masses\footnote{We have chosen the convention to always take
  the lighter particle as first in the pair.}.   

Let us discuss various limits in Eq.~\eqref{eq:avxoutfrombeta}. When
there is no flow ($\beta_f = 0$), all average emission 
points are zero and the asymmetry vanishes. If the temperature is very
large compared to the flow velocity (or more generally the random
component dominates over the correlated one) the average emission
point is zero (center of the source). If this happens for both
particles, the asymmetry is zero. If it happens for only
one of them, the asymmetry exists. If the flow velocity strongly
dominates over  temperature, and particles are emitted from the same
system, both average emission points are strongly shifted by the
same amount and in consequence the asymmetry is again small. From the
discussion we see that the existence of emission asymmetry is not
trivial and only arises in a system where both random(thermal) and
correlated(flow) velocities exist and are comparable in magnitude.


We will illustrate the consequences that this mechanism has for the
particle's emitting regions using the {\tt THERMINATOR} calculations
for central Au+Au collisions at the top RHIC energy as an
example. First, we focus on the ``primordial'' particles, that is
particles coming directly from the hydrodynamical stage. Let us
consider two variables: 
\begin{eqnarray}
x_{out} &=& \vec x \vec p_{T} / p_{T} \nonumber \\
x_{side} &=& \vec x \times \vec p_{T} / p_{T}.
\label{eq:xoutside}
\end{eqnarray}
The first one is the component of the particle's transverse emission
vector parallel to the particle's transverse momentum; the other is the
perpendicular component. In the upper--left panel of 
Fig.~\ref{fig:cankn} the distribution of these emission components 
are shown for primordial kaons. A strong correlation is seen - all
particles moving ``upwards'' (to the positive ``out''
direction) are emitted from the positive $x_{out}$ part of the
source. This focusing of particle's emission has two distinct effects:
the overall size of the emitting region shrinks both in the ``out''
and the ``side'' direction and the average emission position shifts in
the ``out'' direction, but not in ``side''. Looking at the upper--left
panel of Fig.~\ref{fig:canpi} one sees the same effects but to a much
smaller degree for pions with the same {\it velocity}: the size is
shrunk, but not as much, the average emission position is also
shifted but by a smaller amount. The exact numerical values for the
``size'' (or variance) and ``shift'' (or mean) of these distributions
can be found in Tab.~\ref{tab:sizes}. Fig.~\ref{fig:canpr} shows
analogous pictures for protons. As expected, the effect is the
strongest here, as protons have almost twice the mass of kaons. 

From the discussion above, and the numerical values in
Tab.~\ref{tab:sizes} and~\ref{tab:shifts}, one immediately sees that
hydrodynamics produces two distinct trends in femtoscopic observables:
(a) The size of the emitting system gets smaller with increasing
$p_{T}$ of the particle. This effect is well understood theoretically
(so called ``lengths of homogeneity'') and universally observed
experimentally in femtoscopy in heavy-ion collisions (``$m_{T}$
scaling'')~\cite{Lisa:2005dd}. (b) Average emission points of
particles with different $p_{T}$ (for example with same velocity but
different mass) are different, and this difference is well defined:
lighter particles appear to be, on average, emitted closer to the
center of the source\footnote{Note that the values in
Tab~\ref{tab:sizes} are calculated vs. the single-particle momentum
direction, which is only the approximation of the $out$
direction. Values in Tab.~\ref{tab:shifts} are calculated vs. the
proper $out$ direction - the total momentum of the pair. Therefore
small differences between the shift values in the two tables are to be
expected.}.   

The second trend is the main focus of non-identical particle
analysis. It so happens, that the non-identical particle femtoscopy
correlates particles with {\it the same velocity} but possibly with
{\it different masses}. Moreover, it has a unique feature of being able
to measure not only the ``size'' (more precisely - the second moment
of the two-particle distribution - the variance) but also the
``shift'' (i.e. the first moment of the distribution - the mean)
between average emission points. It is therefore able to directly test
the predictions of the hydrodynamic model with respect to the $x-p$
correlations. We would like to emphasize that this is the most direct
and unambiguous signal of collectivity available to femtoscopy. The
$m_{T}$ scaling, which is predicted by hydrodynamics, can also be
explained by other mechanisms not requiring collectivity
(e.g.\ ``temperature gradients''~\cite{Gombeaud:2009fk}); but no
mechanism is known, which would produce such specific emission
asymmetries with no collectivity. 

The asymmetry is predicted to arise only in the ``out'' direction. The
``side'' asymmetry is zero. It can also be shown that for rapidity
symmetric systems of a collider, such as RHIC, when the target is
identical to the projectile, the longitudinal asymmetry is also
expected to vanish. 

\subsection{``No directional correlation'' test}

In the last section we have argued that the correlation between
spatial emission angle $\phi_r$ and particle's velocity direction
$\phi_f$ is responsible for the emission asymmetries. We have
tested this argument by performing a calculation in which we
have intentionally broken this correlation, in order to show that in
such case no asymmetries arise. We take each {\it primordial} particle
separately. From its original transverse emission coordinates $(x_o,
y_o)$ we calculate its transverse emission radius $r$ and angle
$\phi_r$. Then, we randomize the angle $\phi_r$, but keep the emission
radius $r$ unchanged, and calculate the new ``scrambled'' emission
point $(x_s, y_s)$. If the particle is unstable and consequently
decays, the emission points of all daughter particles are shifted by
the same amount, calculated for the parent particle $(x_s - x_o, y_s -
y_o)$. Note that the momentum observables are not affected.

The upper--right panels of Figs.~\ref{fig:cankn},~\ref{fig:canpi} and
\ref{fig:canpr} show the effect of this procedure on primordial
particles. As expected, the average emission point of all particle
types is now at $(0,0)$, and the asymmetry between particles of
different masses, shown in Tab.~\ref{tab:shifts} is zero. Our claims
are confirmed: the model with no directional correlation shows no 
asymmetry. 

\subsection{Importance of resonances}


The Lhyquid+{\tt THERMINATOR} model assumes that the evolution of the
heavy-ion collision proceeds in stages. After the initial
non-equilibrium phase there is a collective phase, which can be well
described by hydrodynamic equations, with the equation of state
assuming the existence of the Quark Gluon Plasma. However, at some
point the system becomes so dilute that the continuous medium
description of hydrodynamics is no longer viable and one converts the
system to hadronic degrees of freedom - the {\tt THERMINATOR} model
does it via the von-Neumann sampling of the probability distribution
obtained from the Cooper-Frye formula. All known resonances are the
degrees of freedom in the hadronic phase. Their abundances are well
described by chemical models, and if one trusts these calculations,
then at least 2/3 of the observed pions do not come from the original
hydrodynamic phase (the so-called ``primordial'' particles), but are
daughters of resonances. Additionally all resonances have their
intrinsic lifetime, so they travel some distance before decaying -
hence they will certainly modify the space--time picture of particle
emission.  It is clear that a careful and detailed simulation of the
resonance propagation and decay (e.g.\ as implemented in {\tt
THERMINATOR}) is a critical feature of the model which aims to
describe femtoscopic observables. Also, the relative abundances and
decay momenta of resonances producing pions, kaons and protons as
final particles are obviously quite different; so, the feature is even
more important for non-identical particle correlations, where we study
relative differences between various particle types.

Let us first qualitatively consider a resonance decay process. We have
the original resonance, produced as a primordial particle. Its
emission followed common ``flow'' $x-p$ correlation, as for any
other particle. Resonances are usually quite heavy, so we expect
this correlation to be strong. The resonance travels some distance
with the original velocity - this enhances the $x-p$
correlation. After some random time the decay process occurs. Two (or
three) particles are  
created at the decay point. We note the similarity between the
resonance decay process and the emission from the fluid element. The
daughters of the resonance will have the ``common'' velocity - this
time it is not the fluid elements velocity, but simply the
resonance's one. And they will have the ``random'' component - the
``decay momentum'' of the given decay channel. And again the random
component will matter more for the lighter particle - the velocity
that corresponds to the fixed ``decay momentum'' will be larger for
the lighter particle. One can imagine two scenarios. If the decay
momentum of the given decay channel is large, compared to the
daughters' mass, than the daughters' emission direction will be
randomized, and the common $x-p$ correlation will be lost. On the
other hand if the decay momentum is small, the correlation will be
preserved (or, in other words, the daughter particle will travel in
roughly the same direction as the parent) or even enhanced due to the
additional $x-p$ correlation resulting from resonance propagation.

From the description above one concludes that the resonance process
will induce the space-momentum correlation in a way similar to the
collective flow; so, one can ask if it can be an alternative mechanism
producing such asymmetries. However, qualitative expectations for
values of the asymmetries or even general trends are not immediately
obvious. They will non-trivially depend on relative abundances of
resonances and their daughters, as well as particular values of decay
momenta in specific decay channels. The fact that some resonances
decay in cascades makes it even more
complicated~\cite{Lednicky:1992me}. Moreover the original 
primordial resonances will have a natural $x-p$ correlation coming
from the earlier hydrodynamic phase. One then faces a quantitative
problem: do the resonance decays introduce emission asymmetries on
their own, independent from flow asymmetries? And if yes, how big are
they, compared to the asymmetries coming from flow? In particular are
they small enough, so that one can still safely interpret the
asymmetry observed in the experiment as coming from collective
behaviour such as flow?

To answer these questions a detailed simulation is needed, in particular
one that has intrinsic $x-p$ correlations coming from flow (preferably
with the possibility to switch them off) and which incorporates all
known resonances, together with the state-of-the art knowledge of
their masses and decay channels. From this description it is clear
that {\tt THERMINATOR} model is perfectly suited for the task. In
addition, we use the ``scrambling'' procedure described in the previous
Section to switch off the $x-p$ correlations coming from flow, in
order to estimate the asymmetry coming from the decay processes alone.

Let us see what is the effect of resonance decays on the properties of
the emitting regions. In panels e) of
Figs.~\ref{fig:canpi},~\ref{fig:cankn},~\ref{fig:canpr} the emission
points of all particles (both primordial and from resonances) are
shown, while Tab.~\ref{tab:sizes} shows the numerical values of sizes
and shifts. The resulting asymmetries are listed in
Tab.~\ref{tab:shifts}. We consider the size of the system first. As
expected the overall size is larger - that is naturally expected as
the resonances travel some distance before they decay. The $m_{T}$
scaling seems to be preserved. In general the trend is consistent with
previous studies of resonance influence on femtoscopic
observables~\cite{Kisiel:2006is,Lednicky:1992me} and agrees with the
natural expectations. Less trivial and more interesting effects are
visible in the average emission points. A qualitatively different
effect is seen for pions than for kaons and protons. We first inspect
the emission points for particles coming from resonances only (no
primordial ones), seen in  panels c) of
Figs.~\ref{fig:canpi},~\ref{fig:cankn},~\ref{fig:canpr}. For pions 
the average emission point is shifted {\it less} from the center for
resonance daughters, than for the primordial ones. Apparently the
first scenario described three paragraphs before is in effect here: the
decay momenta of the resonances producing pions are so large, compared
to the pion mass, that they completely wash out the original flow
$x-p$ correlation. In contrast both for kaons and protons the
resonance daughters are shifted {\it more} from the center than
primordial particles. This time the second scenario is in effect - the
decay  momenta are small compared to particles' masses, so small, in
fact, that they are not even able to counter the additional $x-p$ 
correlation coming from resonance propagation. A more detailed
discussion of this effect, with examples of particular resonances and
decay channels for pions and kaons is found
in~\cite{Kisiel:wpcf08}. The effect persists when one takes all
particles, primordial and resonance daughters, together. Both effects
collaborate in enhancing the pion--kaon and pion--proton asymmetries:   
\begin{eqnarray}
\left < r_{out}^{\pi K} \right > &\approx& \left < x_{out}^{\pi} \right > - \left <
x_{out}^K \right > \nonumber \\ 
\left < r_{out}^{\pi p} \right > &\approx& \left < x_{out}^{\pi} \right > - \left <
x_{out}^p \right >  
\label{eq:asymmdef}
\end{eqnarray}
while the kaon--proton asymmetry stays rather similar to the primordial
only case (and small):
\begin{eqnarray}
\left < r_{out}^{K p} \right > &\approx& \left < x_{out}^{K} \right > - \left <
x_{out}^p \right > .
\label{eq:asymmkpdef}
\end{eqnarray}
We compare the source distributions in panels on the left in
Figs.~\ref{fig:canpi},~\ref{fig:cankn},~\ref{fig:canpr} to the ones in
panels on the right, where the ``scrambling'' procedure was applied. The
numerical values for asymmetries are given in
Tab.~\ref{tab:shifts}. First, let us focus on pion--kaon and pion--proton
pairs. As already discussed, ``scrambled'' primordial particles show
no asymmetry. The non-primordial ones do show some, but still
significantly lower than the non-scrambled ones. This shows that for
non-primordial particles {\it both} sources of asymmetry are important:
the original $x-p$ correlation of the parent particle and the
additional asymmetry from the decay process. However, the former
dominates. A critical test is the comparison of asymmetries for all
particles between normal ``flow'' and ``scrambled'' scenarios, since
this is the observable measured in the experiment. As the simulation
shows, the asymmetry which can be attributed solely to the trivial
resonance decay processes can account for only 20\% (17\%) of the total
asymmetry produced in central Au+Au collisions at top RHIC energy for
the pion--kaon (pion--proton) pair. For kaon--proton the additional
asymmetry produced by the resonance decays is negligible. 

In summary, we have shown that resonance decays do not dilute,
but rather enhance the asymmetry signal for pion--kaon and
pion--proton pairs. At the same time, even though the resonance decay
process can potentially be an independent source of emission
asymmetry, we have shown, by detailed calculations, that in realistic
conditions such asymmetry is less than a quarter of that produced by
flow. Therefore our original expectation holds: if significant
emission asymmetry is observed in pion--kaon and pion--proton
correlations in the experiment, it favors the explanation of
strong $x-p$ correlations in the emitting system, such as the ones
produced by hydrodynamic radial flow. 

\subsection{Flow versus time asymmetries}

\begin{table}[tb]
\caption{Time and space asymmetries (from the fits to the
distributions around their peaks) for central Au+Au collisions. PRF
values are with asterisk.  
\label{tab:timespace}}
\begin{tabular}{|r|cc|cc|c|}
\hline
\hline
 & $\left < r_{out} \right >$~~~ & $\left < \gamma_t \right > \left <
r_{out} \right >$ & $\left < \Delta t \right >$~~~ & $-\left <
\beta_t \right > \left < \gamma_t \right > \left < \Delta t
\right >$ & $\left < r^{*}_{out} \right >$ \\ 
\hline
$\pi K$ all & -3.3 & -5.0 & 2.7 & -3.0 & -8.0 \\
     ~~~pri & -1.6 & -2.4 & 1.5 & -1.7 & -4.1 \\
\hline
$\pi p$ all & -4.0 & -5.7 & 3.8 & -3.5 & -9.2 \\
     ~~~pri & -2.4 & -3.4 & 2.1 & -2.0 & -5.4 \\
\hline
$K p$ all & -0.8 & -1.1 & 0.9 & -0.6 & -1.7 \\
   ~~~pri & -1.0 & -1.3 & 0.5 & -0.4 & -1.7 \\
\hline
\end{tabular}
\end{table}

\begin{figure}[tb]
\begin{center}
\includegraphics[angle=0,width=0.4 \textwidth]{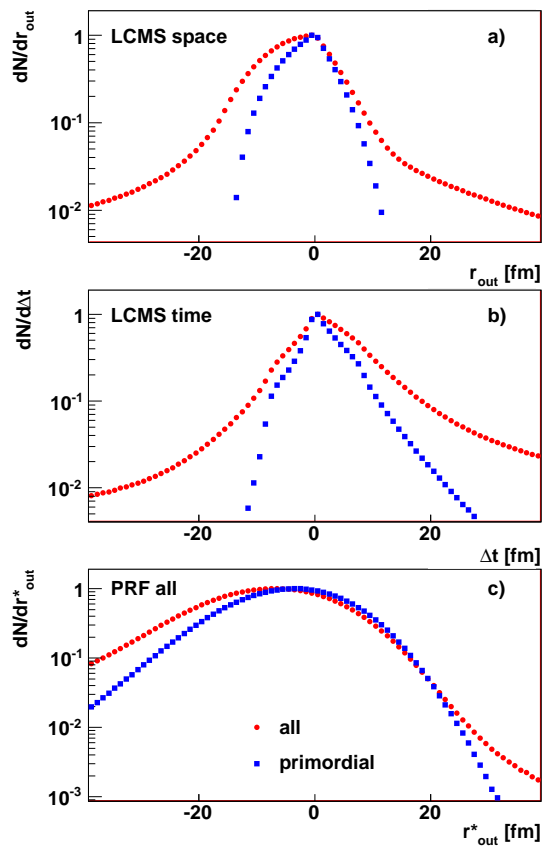}
\end{center}
\vspace{-6.5mm}
\caption{(Color on-line) Pion-kaon emission functions in LCMS for
  space a) and time b) components, combined into the observable
  asymmetry in PRF c). Red circles are for all particles, blue squares
  for primordial only.   
\label{fig:canbeta}}
\end{figure}

As we have noted in the Introduction, the non-identical particle
correlations technique has been initially developed to measure
mainly the emission time differences between various particle
species. Only later the connection to radial flow was
noted~\cite{Lednicky:1999nn,Lednicky:2003fe}.   

In Au+Au collisions one expects intrinsic time differences for at
least two reasons. The reader is referred to~\cite{Kisiel:2008ws},
where the evolution of the average emission times with particle's
$p_T$ was discussed. It was noted that due to the particular features
of the hydrodynamic emission function, particles with larger $p_T$
were, on average, emitted earlier. In addition, studies with the
``blast-wave'' parametrization~\cite{Retiere:2003kf} showed that a
time difference between particles of different masses arises in
boost-invariant models (the influence of the violation of boost
invariance on the time and longitudinal shifts has been discussed
in~\cite{Lednicky:xderiv}). In our case both effects would contribute
to the additional asymmetry in PRF: 
\begin{equation}
r^{*}_{out} = \gamma_t \left (r_{out} - \beta_t \Delta t \right ),
\label{eq:rsfromrt}
\end{equation}
which would go {\it in the same direction} as the spatial one coming
from radial flow. Even though it cannot be directly correlated with
the radial flow, it is still very much hydrodynamic in nature and we
do not consider it as an alternative, non-collective explanation of
the asymmetry.  

However, the resonance decay process can also introduce additional
time asymmetries, as resonance decays occur with a certain time delay
and if some particles are more abundantly produced by resonances than
others, asymmetries may arise. In contrast to the effect discussed in
the previous paragraph, these time delays are not ``hydrodynamic'' in
nature and should be treated as alternative sources of
asymmetry. Moreover, one expects pions to be most abundantly produced
by resonance decays, which would mean that they would appear to be, on
average, produced later than kaons and protons, producing asymmetry
{\it in the same direction} as the flow. Clearly the matter requires
careful quantitative study. 

In Fig.~\ref{fig:canbeta} the time and space emission asymmetries are
shown on the left and central panels. One sees that both are present
and are significant. Resonance decays modify both distributions and
add a complication of long-range tails. Therefore, a simple
mean of a distribution is no longer a good variable to characterize
such asymmetries. One is forced to use the mean values of the
functional forms fitted to the distributions around the
peak. Tab.~\ref{tab:timespace} shows the summary of these fits. We 
see that even for the primordial particles only, there is already a
time difference, although smaller than the space one. Introducing
resonances increases the time asymmetry. Evidently the expected effect
is seen. Comparing the values in PRF we take the difference
between ``all'' and ``primordial'' cases for the time asymmetry as an
estimate on how much time asymmetry the resonance decays
introduce. Both for pion--kaon and pion--proton the ``non-hydrodynamic''
asymmetry coming purely from resonance time delay is less than 15\% of
the predicted overall asymmetry. For kaon--proton it is less than
25\%. Again, the flow asymmetry dominates the calculated asymmetry
signal. 

\subsection{Expectations for qualitative trends}

Following the discussion in the previous subsections one can
formulate several predictions of expected qualitative trends. We will
consider pions, kaons and protons in specific $p_{T}$ windows,
corresponding to the acceptance of the STAR experiment at RHIC. This
means that the velocities of these particles will be 
fixed, and one might assume that the same pions will be correlated
with kaons for the pion--kaon correlation and with the protons for the
pion--proton correlation. Since we are correlating particles with
similar velocity, we will be correlating very low-$p_{T}$
($\approx~0.1$~GeV/c) pions with medium-$p_{T}$ kaons
($\approx~0.5$~GeV/c) and moderate-$p_{T}$ protons
($\approx1$~GeV/c). We give a detailed relation between
single-particle and two-particle sizes in
Appendix~\ref{sec:emfuncs}. In terms of the observables themselves,
that is the two particle variances, one expects that $R_{\pi K}$
and $R_{\pi p}$ will be similar and large, since they are
dominated by the large low-$p_{T}$ pion size. In contrast $R_{K
p}$ is expected to be significantly smaller. 

\begin{figure}[tb]
\begin{center}
\includegraphics[angle=0,width=0.4 \textwidth]{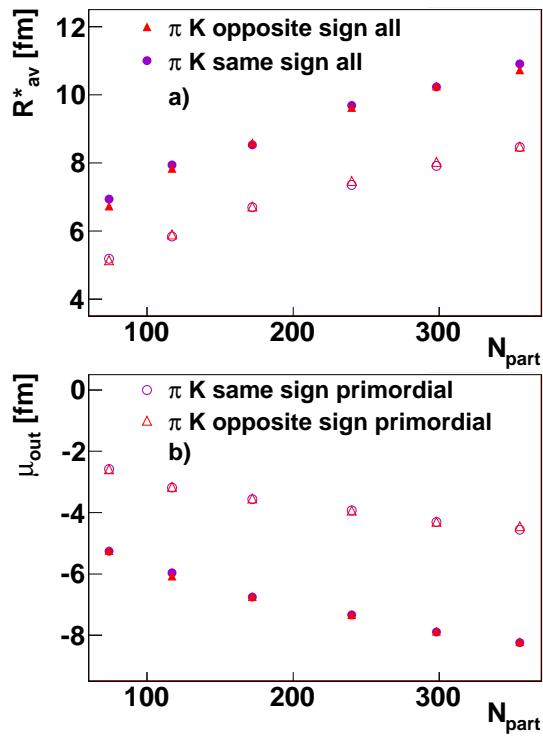}
\end{center}
\vspace{-6.5mm}
\caption{(Color on-line) The pion--kaon overall radius a) and
emission asymmetry b) in PRF as a function of centrality for Au+Au
collisions at $\sqrt{s_{NN}}=200$~GeV. Close points are for all
particles, open for primordial only. Circles are same-sign pairs,
triangles - opposite-sign. 
\label{fig:pikdep}}
\end{figure}

The asymmetries show a common feature - with the definition of
Eqs.~\eqref{eq:asymmdef},\eqref{eq:asymmkpdef} (the lighter particle always
taken as first) they are all negative - which reflects the fact that
lighter particles are expected to be emitted closer to the center of
the system than the heavier ones. 

\begin{figure}[tb]
\begin{center}
\includegraphics[angle=0,width=0.4 \textwidth]{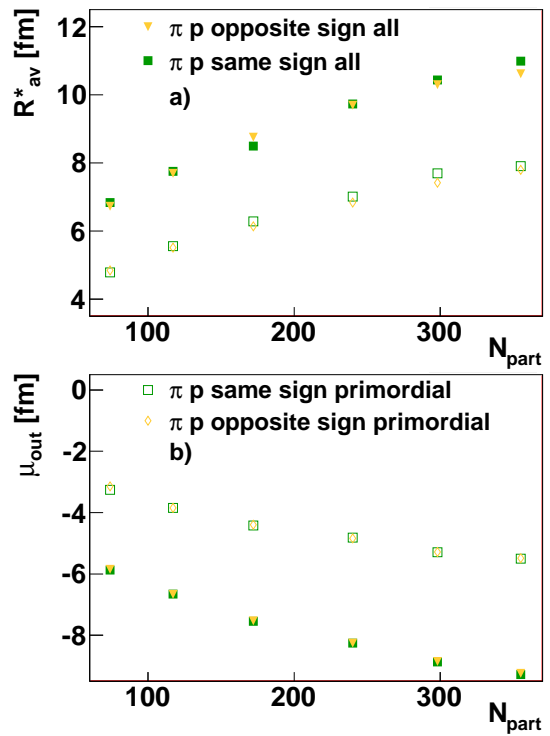}
\end{center}
\vspace{-6.5mm}
\caption{(Color on-line) The pion--proton overall radius a) and
emission asymmetry b) in PRF as a function of centrality. Close
points are for all particles, open for primordial only. Squares are
same-sign pairs, triangles (diamonds) - opposite-sign.
\label{fig:pipdep}}
\end{figure}

As for the relations between asymmetries, hydrodynamics naturally
predicts that the $\mu_{\pi p}$ will be the largest, $\mu_{\pi K}$
will be of similar magnitude but smaller, while $\mu_{K p}$ will be
much smaller than the other two. It can also be shown that the
following relation should hold:

\begin{equation}
\mu_{\pi p} = \mu_{\pi K} + \mu_{K p}
\label{eq:msumrule}
\end{equation}

The asymmetries for ``all'' particles for pion--kaon and pion--proton
pairs should be significantly different than for ``primordial'' only,
while for kaon--proton the difference should be small.

\begin{figure}[tb]
\begin{center}
\includegraphics[angle=0,width=0.4 \textwidth]{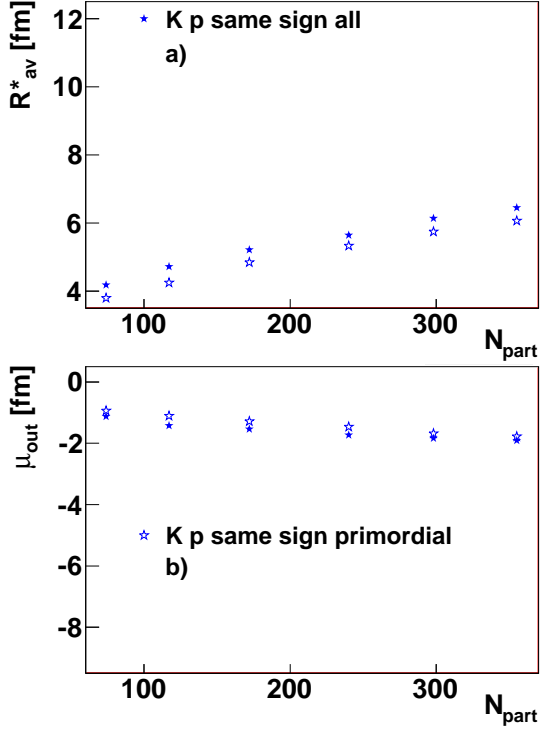}
\end{center}
\vspace{-6.5mm}
\caption{(Color on-line) The kaon--proton overall radius a) and
emission asymmetry b) in PRF as a function of centrality. Close
points are for all particles, open for primordial only.
\label{fig:kpdep}}
\end{figure}

In Figs.~\ref{fig:pikdep},~\ref{fig:pipdep},~\ref{fig:kpdep} the
predicted system size and emission asymmetry for pion--kaon,
pion--proton and kaon--proton pairs respectively, calculated by
Lhyquid+{\tt THERMINATOR}, are shown. All the expected trends
mentioned in this section are confirmed.

\subsection{Centrality dependence}

We first consider the centrality dependence of the system size. The
assumed initial conditions have a clear dependence: the system size
and initial temperature grow with $N_{part}$; both should result in
larger sizes thorough of the evolution. The hydrodynamic evolution
assumes identical equation of state and identical freeze-out
temperature for all centralities; so, there is no reason to 
expect that this dependence will be altered in the final state. Also,
calculations for identical pions show the same trend. Finally,
all available experimental data, including pion, kaon and proton
femtoscopy show the same trend. Therefore increase of the system size
with $N_{part}$ is expected.

\begin{figure}[tb]
\begin{center}
\includegraphics[angle=0,width=0.4 \textwidth]{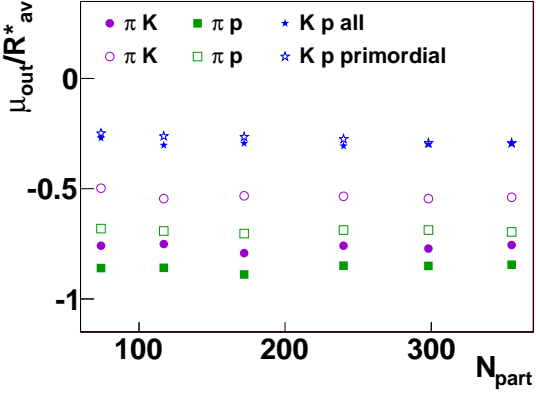}
\end{center}
\vspace{-6.5mm}
\caption{(Color on-line) The emission asymmetry scaled by the overall
system size in PRF. Closed points are for all particles, open for
primordial only. Circles are for pion--kaon pairs, squares for
pion--proton, stars for kaon--proton.
\label{fig:cankprat}}
\end{figure}

The asymmetry is the result of a hydrodynamic evolution. As already
mentioned the parameters of the evolution do not change with
centrality, only the initial conditions. Since the chemical properties
of matter do not depend on centrality either, the resonance decay
phase is not expected to be very different, giving an enlargement of
the system size by a constant amount. On the other hand, as the
collisions become more peripheral, the initial overlap region is
shrinking, so one expects that the long-range resonance corona will
become relatively more important. Since we know that the overall size
of the system grows with centrality, the absolute value of asymmetry
will also grow. The ratio of asymmetry to the system size is shown in 
Fig.~\ref{fig:cankprat} and indeed we can see that the scaling of
$\mu$ with $R^{*}_{av}$ holds well for all centralities, all pair
types, both for ``all'' and ``primordial only'' particles. 

\section{Non-identical particle femtoscopy formalism}
\label{sec:nonidform}

In femtoscopy one aims to measure the space--time configuration of the
emission process in hadronic collisions by analyzing the specific
behavior of the two-particle correlation function. A natural variable
versus which this correlation is measured is the half of generalized
pair relative momentum $k^{*}$, as opposed to analyses focused on
event structures, which  use e.g.\ azimuthal angle and pseudorapidity
differences. $k^{*}$ is calculated in the pair rest frame, so it is
also the momentum of the ``first'' particle of the pair in the
PRF. The particles in the pair are different, so it is important to
define which one is ``first''. Later we give the conventions which we
use in this paper. Femtoscopy also requires a precise knowledge of
particle type, which means experiments wishing to do such analysis
must have good particle identification capability. 

The correlation function is defined as:
\begin{equation}
C(\vec p_a , \vec p_b ) = \frac {P_2(\vec p_a, \vec p_b)} {P_1(\vec
p_a) P_1(\vec p_b)},
\label{eq:cfgeneral}
\end{equation}
where $P_2$ is a conditional probability to observe a particle with
momentum $\vec p_b$ if a particle of momentum $\vec p_a$ was also
observed, while $P_1$ is a simple probability to observe a particle
with a given momentum. Note that this definition is general and
applies equally well to the femtoscopic correlation function and to
other two particle correlations e.g.\ event structure ones. 

The experimental femtoscopic correlation function is a specific form
of Eq.~\eqref{eq:cfgeneral}, written as:
\begin{eqnarray}
C(\vec k^{*}) &=& \frac {\int A(\vec p_a, \vec p_b) \delta (\vec k^{*} -
\frac {1}{2} (\vec p^{*}_a - \vec p^{*}_b)) d^3 p_a d^3 p_b } 
{\int B(\vec p_a, \vec p_b) \delta (\vec k^{*} -
\frac {1}{2} (\vec p^{*}_a - \vec p^{*}_b)) d^3 p_a d^3 p_b} \nonumber \\
& \equiv & \frac {A(\vec k^{*})} {B(\vec k^{*})} ,
\label{eq:cfexp}
\end{eqnarray}
where $A(\vec p_a, \vec p_b)$ is the distribution of correlated pairs
(i.e. both particles coming from the same event) of particles of type
$a$ and $b$ and $B(\vec p_a, \vec p_b)$ is the same distribution, but
particles are not correlated (i.e. come from two different
events). Note that the argument of $C$ is changed to half of the pair
relative momentum $k^{*}$. Also note that with this definition the
correlation function will contain not only femtoscopic 
correlations, but also all other event-wide correlations projected to
the two-particle space. These include: elliptic flow $v_2$, global
event energy and momentum conservation, resonance decay correlations
(if $a$ and $b$ are different and can be products of the decay of a
given resonance, e.g.\ products of the $\Delta^{++}$ resonance in the 
$\pi^{+}$ -- $p$ correlation function), residual correlations
(remnants of the femtoscopic correlations between parent particles,
which decayed weakly into the particles of interest, e.g.\ residual
correlations between $\Lambda$ and $p$ feeding into the $p$--$p$
correlations), jets, etc. There are numerous experimentalists recipes
to construct $B$ in such a way that these correlations are included
there. In that case, dividing $A$ and $B$ also divides out the
correlations. One should also correct for other non-femtoscopic
effects, so that one is left with pure femtoscopic correlation in $C$.
The latter is desirable, since it is the femtoscopic-only effect that is
usually computed in models. But such correction procedures are never
fully effective; so, one must take it into account when comparing the
``compound'' correlation functions from the experiment with the
``pure'' femtoscopic CF from models. The details of such procedures
are clearly experiment dependent and are beyond the scope of
this paper. We only note that some models may include the effects of
global correlations (e.g.\ elliptic flow, energy-momentum conservation);
so, they can be used to model such effects. We discuss how this can be
done later in the chapter. 

In models one defines the correlation function via the single and
two-particle emission functions:
\begin{eqnarray}
S_{A}(\vec x_1, \vec p_1) & = & \nonumber \\
\int S(x_1, p_1, x_2, p_2, ..., x_N, p_N) dx_2 dp_2 ... dx_N dp_N & & \\
S_{AB}(\vec x_1, \vec p_1, \vec x_2, \vec p_2) &=& \nonumber \\
\int S(x_1, p_1, x_2, p_2, ..., x_N, p_N) dx_3 dp_3 ... dx_N dp_N & &,
\label{eq:sffromall}
\end{eqnarray}
which are interpreted as a probability to emit a particle (a pair of
particles) from a given space--time point with a given momentum. With
this definition one might substitute $S_{AB}$ for $P_2$ and $S_{A}$
for $P_1$ in~\eqref{eq:cfgeneral}. Such a correlation function is a
14-dimensional object (7 independent components per particle: 4
space--time and 3 momenta). Assumptions are needed to reduce the
number of dimensions. We describe them below.

In principle, the source emission function $S_{AB}$ should reflect all
the physics aspects of the particle emission process, including the
proper symmetrization for bosons and fermions, as well as the
influence of the two-body and many-body Final State
Interaction (FSI). However, commonly used heavy-ion collision models do not
include these effects. Instead, one assumes that each particle's emission
process is independent, or in other words that an interaction between
the two final-state particles after they are created is independent
from their emission process. The introduction of this factorization of
the FSI and two-particle wave function symmetrization gives the
equation:  
\begin{eqnarray}
C(\vec p_1, \vec p_2) = \int S_{AB}(\vec p_1, {\bf x_1}, \vec p_2,
{\bf x_2}) \left | \Psi_{AB} \right |^2 d^{4} {\bf x_1} d^4 {\bf x_2}  ,
\label{eq:KonPrat}
\end{eqnarray}
where $\Psi$ is the pair wave function. Particle types $A$ and
$B$ are known, so the momenta have only three independent components,
while for positions all four components are 
independent. We also mention that the equation~\eqref{eq:KonPrat} has
strong similarities to the Fermi equation used to describe the $\beta$
decay process~\cite{Lednicky:2007ax}.

\subsection{Pair wave function}

The pair wave function $\Psi$ describes the behavior of a pair of
particles, one of type $A$ and another of $B$. In non-identical particle correlations
we use a particular form of $\Psi$, which corresponds to the following
physical scenario: two particles $A$ and $B$, which, shortly
after they are produced in heavy-ion collision, interact via the
FSI, in our case Coulomb and/or
Strong. After this interaction they propagate to the detector as
plane-waves. In that case, the particular form of $\Psi$ is the solution
of the scattering problem, viewed with the reversed time direction. We
also use equal time approximation which assumes the particles were
born at the same time in PRF
(see~\cite{Lednicky:2005tb,0954-3899-35-12-125109} for a detailed
description  
and estimation of the systematic error introduced by such
assumption). We factorize the wave function into the
part describing the motion of the pair as a whole (a function of pair
total momentum and ``average'' emission point), and the component
describing the interaction itself -- dependent on pair relative
momentum $\vec k^{*}$ and separation $\vec r^{*}$. The first component produces
only an additional phase, which does not influence the modulus of 
the wave function. Since in our study we are only interested in the
modulus, we can neglect this component, and we are left
with~\cite{Lednicky:2005tb}:  
\begin{equation}
\Psi^{(+)}_{-k^{*}}({\bf r^{*}}) = \sqrt{A_C (\eta)} \left [ e^{-i
\vec k^{*} {\bf r^{*}}} F(-i \eta, 1, i \zeta) + f_C (\vec k^{*})
\frac {\tilde G (\rho, \eta)} {{\bf r^{*}}} \right ]
\label{eq:fullpsi}
\end{equation}
where $A_C$ is the Gamow factor, $\zeta = k^{*} r^{*} (1 +
\cos{\theta^{*}})$, $\eta = 1/(k^{*} a_C)$, $F$ is the confluent
hypergeometric function, $\tilde G$ is the combination of the regular
and singular s-wave Coulomb functions and $f_C$ is the strong
scattering amplitude, modified by the Coulomb
interaction. $\theta^{*}$ is the angle between the pair relative
momentum $k^{*}$ and relative position $r^{*}$ in PRF, while $a_C$ is 
the Bohr radius of the pair, equal to  $248.52$~fm, $222.56$~fm and
$83.59$~fm for pion--kaon, pion--proton and kaon--proton pairs
respectively, and is negative for opposite-charged pairs. For
identical particles $\Psi$ must also be properly
(anti-)symmetrized. This equation is valid in the outer regions of the
Strong Interaction potential, and neglects the components for angular
momentum $l \geq 1$, the latter is a valid approximation for small
$k^{*}$. 

In femtoscopic analysis we assume that we 
know $\Psi$ with infinite accuracy, so that we can try to invert
Eq.~\eqref{eq:KonPrat} to obtain, from the measured correlation
function $C$, the information about the emission function $S_{AB}$.
In this work we focus on non-identical combinations of the most abundant
stable hadrons measured in heavy-ion collisions: pions, kaons and
protons. For each of the combinations (pion--kaon, pion--proton and
kaon--proton) there are four charge combinations: two of the same sign and
two of the opposite sign. The wave function for
both same-sign (and both opposite-sign) combinations are identical. We
note that the wave function characteristics are indeed well known for
all combinations, except for the opposite-sign kaon--proton. The strong
interaction in this system is interesting in its own right, and is the
focus of intense theoretical investigation (see
e.g.~\cite{Borasoy:2005ie}). The femtoscopy technique can  
be useful in this regard -- by inverting the problem and assuming that
we know the source distribution (from other femtoscopic measurements)
we can invert Eq.~\eqref{eq:KonPrat} to deduce the parameters of
$\Psi$ from the measured correlation function. Similar technique can
also be used to study the strong interaction in the $\pi^{+} \pi^{-}$
system precisely. Such studies are beyond the scope of this paper
(see, e.g.,~\cite{Lednicky:2005tb,Lednicky:xderiv}).

For the systems analyzed in this work, $f_C$ can be parametrized
in the effective range approximation by:
\begin{equation}
f_C(\vec k^{*}) = \left [ \frac {1} {f_0} + \frac {1} {2} d_0 \vec
{k^{*}}^2 - \frac {2} {a_C} h(\vec k^{*} a_C) - i \vec k^{*} A_C (\vec
k^{*}) \right ]^{-1} ,
\end{equation}
where the $f_0$ is $0.137$~fm for same-sign pion--kaon pair,
$-0.071$~fm for opposite-sign pion--kaon pair, $-0.148$~fm for
same-sign pion--proton, $0.112$~fm for opposite-sign pion--proton,
$-0.360$~fm for same-sign kaon--proton pair. The effective radius
$d_0$ can be put equal to $0$ for all considered pairs, at small
$k^{*}$ where the $1/f_0$ term dominates.  

The full form of the wave function~\eqref{eq:fullpsi} must be used
when calculating correlation functions to be compared with 
data. However, for theoretical calculations it is sometimes instructive
(and faster numerically) to neglect the strong interaction. This is
acceptable since the strong interaction is expected to be small for
the pairs of interest, except for opposite-sign kaon--proton pairs,
which we will not include in the model calculations. For simplicity we
will use the Coulomb-only wave function in the discussion in the next
paragraph, but the conclusions hold for full Coulomb+strong wave
function as well.  

For the discussion it is important to provide the form of the $F$
function explicitly: 
\begin{equation}
F(\alpha, 1, z) = 1+ \alpha z + \alpha(\alpha+1)z^2/2!^{2} + ...
\label{eq:ffun}
\end{equation}

\subsection{Emission function}

The FSI correlation, described by~\eqref{eq:fullpsi}, depends only on
relative momentum $\vec k^{*}$ and separation ${\bf r^{*}}$ of the
pair (the angle between the two vectors is $\theta^{*}$). The first
simplification of $S_{AB}$ from~\eqref{eq:sffromall} is to change to
the relative variables and integrate out the sum ones:
\begin{eqnarray}
S_{AB}(\vec k^{*}, {\bf r^{*}})  & =  &\int \int_{P_{min}}^{P_{max}}
S_{AB}({\bf x_1}, \vec p_1, {\bf x_2}, \vec p_2)\nonumber \\
& & \times  \delta (\vec k^{*} -
1/2(\vec p^{*}_1 - \vec p^{*}_2)) \nonumber \\	
& & d^3 p_1 d^3 p_2 \nonumber \\
& &\times  \delta (\vec r^{*} -
(\vec x^{*}_1 - \vec x^{*}_2)) \nonumber \\	
& & d^4 x_1 d^4 x_2
\label{eq:sepsum}
\end{eqnarray}
The integration over space is done over the full variable range. On
the other hand, particles' momenta are measurable, so it is possible to
define the boundary momenta $P_{min}$ and $P_{max}$ for which the
integration is done. In fact, for identical particle femtoscopy it is
common to define several $k_{T}=1/2(p_{T}^1 + p_{T}^2)$ ranges
and create separate correlation functions for each of them. In this way,
the information about the $P$ dependence of $S_{AB}$ is not completely
integrated out and can still be inferred. Up to now similar binning
was not possible for non-identical particle correlations at RHIC,
because of small statistics and limited $p_{T}$ acceptance with good
PID coverage. However, when one compares the model and the experiment
data, one must take care to restrict the $P$ integration range at
least to the one dictated by the $p_{T}$ acceptance of the experiment. 

The emission function~\eqref{eq:sepsum} is a 7-dimensional object. We
use it to rewrite the equation~\eqref{eq:KonPrat},
putting the specific form of the wave function~\eqref{eq:fullpsi} as
well: 
\begin{equation}
C(\vec k^{*}) = \int S_{AB}(\vec k^{*}, {\bf r^{*}}) \left |
\Psi^{(+),AB}_{-k^{*}}({\bf r^{*}}) \right |^2 d^{4} {\bf r^{*}}  .
\label{eq:KonPratRel}
\end{equation} 
This form can be explicitly used to calculate the correlation function
from models. However, it is very rare for a model to provide the full
two-particle emission function. The existing models fall into two
categories: they either provide an analytic form of a {\it single
particle} emission function, or they provide information only about
the produced particles. 

In the first case we assume that each particle's emission process
is independent. Then, the two-particle emission function $S_{AB}$ can
be constructed from single particle emission functions via a
convolution:  
\begin{eqnarray}
S_{AB}(\vec k^{*}, {\bf r^{*}}) &=& \int S_{A}(\vec p_1, {\bf x_1})
S_{B} (\vec p_2, {\bf x_{2}}) \nonumber \\ 
& & \times \delta (\vec k^{*} - \frac {1} {2} (\vec p_{1} + \vec
p_{2})) \delta(r^{*} - {\bf x_{1}} + {\bf x_{2}}) \nonumber \\
& & \times d^{4} {\bf x_{1}} d^{4} {\bf x_{2}} d^{3} \vec p_1 d^{3}
\vec p_2 .
\label{eq:sfromconv}
\end{eqnarray}
In case of identical particles ($A \equiv B$) several simplifications
can be made. The convolution of two identical Gaussians is also a
Gaussian with $\sigma$ multiplied by $\sqrt{2}$. Femtoscopy can
provide information about the two-particle emission function only, but
with the simplifying assumption above, the $\sigma$ of the single
particle distribution can be inferred. For non-identical particles, $A
\neq B$, the above simplified method of comparison is not
applicable. The comparisons with models is more complicated, but
retrieving single-particle source sizes is still possible, provided a
complete set of measurements is performed. The formula is used in
Sec.~\ref{sec:experimental} and derived in
Appendix~\ref{sec:emfuncs}. Also, since generally $S_A \neq S_B$, the  
$S_{AB}$ can yield a non-zero mean value of the separation vector
$\left < \vec r^{*} \right >$. Later in this Section
we explain how this mean value can be accessed experimentally and
argue that this is an important and unique piece of information
accessible via non-identical particle femtoscopy only. 

In the second case, of a model discretely producing particles,
Eq.~\eqref{eq:KonPratRel} is evaluated via the Monte-Carlo procedure: 
\begin{equation}
C(\vec k^{*}) = \frac {\sum_{pairs} \delta(\vec k^{*}_{pair} - \vec
k^{*}) \left | \Psi^{(+),AB}_{-k^{*}}({\bf r^{*}}) \right |^2 } {
\sum_{pairs} \delta(\vec k^{*}_{pair} - \vec k^{*})} .
\label{eq:KonPrarMC}
\end{equation}
Note that if the particles from the model are produced in a correlated
way (e.g.\ with energy and momentum conservation for the full event,
with energy and momentum conservation for resonance decay, etc.)
these are {\it not} destroyed and are still present in $C$. In other
words, this method does not require the simplifying assumption of
Eq.~\eqref{eq:sfromconv}. In this work we use this method to calculate
model correlation functions. Its practical implementation is described
in Sec.~\ref{sec:implement}.

\subsection{Correlation function and asymmetry signal}

\begin{figure}[tb]
\begin{center}
\includegraphics[angle=0,width=0.4 \textwidth]{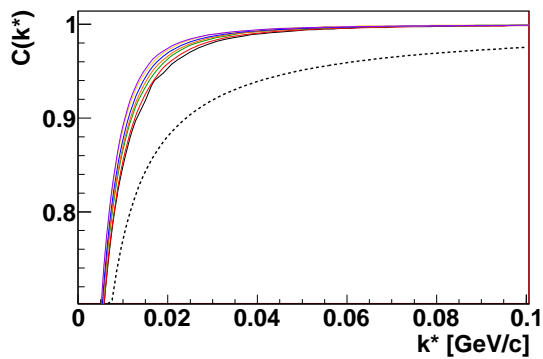}
\end{center}
\vspace{-6.5mm}
\caption{(Color on-line) An example of the non-identical particle
correlation function dependence on the size of the system. Solid lines
represent same-sign pion--kaon correlation functions for a Gaussian
source with sigma from $7.0$~fm (lowest) to $13$~fm (highest) and a
shift of $-5$~fm. The Gamow factor is shown as a dashed line for
comparison. 
\label{fig:pikcfdep}}
\end{figure}

Eq.~\eqref{eq:KonPrat} essentially defines the correlation function as
a pair wave function averaged over the source. Using the simplified
wave function containing only the Coulomb part one can write: 
\begin{equation}
C(k^{*}) = A_{C}(\eta) [1+2 \left < r^{*} (1+\cos \theta^{*}) \right >
/a_c + ...],
\label{eq:capprox}
\end{equation}
where averaging is done over pairs. In this example we illustrate the
origins of the asymmetry effect by considering same-sign pion--kaon pairs. In
such case, $a_c$ is positive, $A_C$ is negative and $r^{*}(1+\cos
\theta^{*})$ is by definition always positive. For a point source
($r^{*} = 0$) the overall correlation effect $R = \left | C - 1 \right
|$ would be maximum and equal to $1-A_C$. For our pairs $A_C-1$ is
negative, while $2 \left < r^{*} (1+\cos \theta^{*}) \right > /a_c$ is
positive, so the two compete with each other. As the size of the
system grows, so does average $r^{*}$, and the correlation effect $R$
decreases. Therefore, the correlation function is sensitive to the
source size; so, we expect to be able to measure the size
of the system. One has to remember that as the size grows, the CF gets
less and less sensitive to the system size; therefore, the analysis is
able to accurately measure only sizes that are not too large. To
determine whether the ``too large'' is larger 
than the maximum expected system size at RHIC is one of the objectives
of this paper. Figure~\ref{fig:pikcfdep} shows an example of how a
non-identical particle correlation function (in this case same-sign 
pion--kaon) depends on the size of the system. One can see that for
a reasonable system size (comparable to the maximum femtoscopic sizes
obtained in the central Au+Au collisions at RHIC) there is a
noticeable and monotonic dependence on the size, indicating that a
qualitative femtoscopic analysis should be possible.

As mentioned in the previous paragraph, the two-particle emission
function can yield a non-zero mean value of the separation vector
$\left < {\bf r^{*}} \right >$. We now discuss 
how it can be observed in data. From Eq.~\eqref{eq:capprox} the
correlation function depends on the angle $\theta^{*}$, between
relative momentum $\vec k^{*}$ and relative position $\vec
r^{*}$. When the two are aligned ($\cos \theta^{*} > 
0$), the correlation effect is smaller, than when they are 
anti-aligned ($\cos \theta^{*} < 0$). The former configuration 
means that the particles of the pair, when born, immediately start to
fly away from each other, so their effective interaction time is
shorter. In the latter case, when they start to fly
towards each other, pass close to each other and only later fly
away. Angle $\theta^{*}$ is not accessible experimentally, but it does
influence $R$.  

Particles' momenta, measured experimentally, can be used to calculate
the relative momentum  $\vec k^{*}$ and pair total momentum $\vec K$,
corresponding to the velocity $\vec v$. The angle between the two is
$\psi$. One divides the observed pairs in two groups: one having $\vec
k^{*}$ and $\vec v$ aligned ($\cos \psi > 0$) and another, having
$\vec k^{*}$ and $\vec v$ anti-aligned ($\cos \psi 
< 0$), and creates two correlation functions, $C_{+}$ and
$C_{-}$. If in the pair sample used to calculate the $C_{+}$ we
have a majority of pairs which also have $\cos \theta^{*} > 0$ (and in
the $C_{-}$ sample the majority of pairs have $\cos \theta^* <
0$),
than $C_{+}$ and $C_{-}$ would differ. In this particular case $C_{+}$ would show
a smaller correlation effect and $C_{-}$ a larger one. When plotting a
``double-ratio'' $C_{+}/C_{-}$ one would see a signal deviating from
unity. For same-sign pairs it would go above unity, while for
opposite-sign pairs it would go below unity.
 
If we see a non-zero ``double-ratio'', it means that $\cos
\theta^{*}$ is somehow correlated with $\cos \psi$. They are connected
via a third angle of interest: the angle $\phi$ 
between the pair velocity $\vec v$ and the pair relative position
$\vec r^{*}$. When we consider only
the projections of all these angles on the transverse plane, we have
trivially:  
\begin{equation}
\psi = \theta^{*} + \phi .
\label{eq:pairangles}
\end{equation}
For the average cosines of these angles, we can write:
\begin{equation}
\left < \cos \psi \right > = \left < \cos \theta^{*} \cos \phi \right > .
\label{eq:cosines}
\end{equation}
By definition for $C_{+}$ the left-hand side of Eq.\eqref{eq:cosines}
is positive. We assumed that $C_{+}$ shows larger correlation effect,
so $\left < \cos \theta^{*} \right >$ is negative. The only way
that the equation can be fulfilled is to also have $\cos \phi < 0$. In
other words it is required that, on average, $\vec r^{*}$ is
anti-aligned with $\vec v$. This is a crucial point, so let us restate
it. If we see a  non-unity double ratio we can conclude that the
average relative position direction is correlated with the pair
velocity direction. That means that we can access, via a rather
straightforward procedure, the mean of the two-particle separation
distribution, which is allowed to be non-zero for non-identical
particles. The emission asymmetry $\left < {\bf r^{*}} \right >$ is a
three-vector, while the consideration above only mentions a single
direction (the direction of the pair velocity $\vec v$). The argument is more
general: we can replace $\vec v$ by any other direction, defined in the
Pair Rest Frame, and repeat the argument to obtain the same
conclusions. The simplest generalization, which we have shown to have
important physics motivation, is the decomposition of $\vec v$ into
components: the longitudinal ``long'' (along the beam axis) and
transverse ``out'' (perpendicular to the beam axis). By performing the
asymmetry analysis versus these two directions we can obtain information
about the ``out'' asymmetry $\mu_{out} = \left < r^{*}_{out} \right >$
and the ``long'' asymmetry $\mu_{long} = \left < r^{*}_{long} \right
>$. For completeness we also use the third direction: ``sidewards'' or
``side'', perpendicular to the other two to study the ``side''
asymmetry $\mu_{side} = \left < r^{*}_{side} \right >$. Each of them
carries important physics information or is useful as an experimental
cross-check; the details will be discussed in the following Sections. 

\begin{figure}[tb]
\begin{center}
\includegraphics[angle=0,width=0.4 \textwidth]{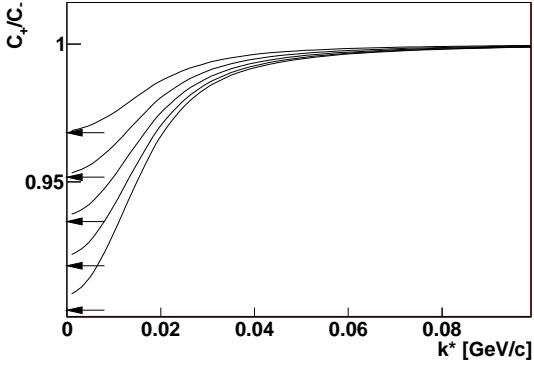}
\end{center}
\vspace{-6.5mm}
\caption{An example of the non-identical particle ``double-ratio''
dependence on the emission asymmetry. Solid lines represent same-sign
pion--kaon ``double-ratios'' for a Gaussian source with sigma of
$10.0$~fm and the asymmetry of $-12.0$~fm (lowest) to $-4$~fm 
(highest). The arrows show the asymptotic values (see text for
details). 
\label{fig:pikdrdep}}
\end{figure}
  
We discussed the possible physical origins of the emission point
asymmetry and its significance in
Section~\ref{sec:modelpreditions}. Now we discuss the general 
properties of the ``double-ratio'' observable. It is allowed to go
both above and below unity, which means either positive or 
negative emission asymmetry (the average emission separation being aligned
or anti-aligned with the velocity direction). It can also be unity,
meaning no emission asymmetry. More detailed analysis also shows that,
for a fixed source size, introducing larger and larger emission
asymmetry produces larger and larger deviation from unity of the
``double-ratio'' signal. This means that one can determine the
existence of emission asymmetry and also measure its magnitude. The
technical details of how it is done are presented in the next
Section. The illustration of the ``double-ratio'' behavior, for a
fixed system size for same-sign pion--kaon pair, is shown in
Fig.~\ref{fig:pikdrdep}.   

It is instructive to derive the equation for asymptotic behavior of
the double ratio $C_{+}/C_{-}$ at $k^{*}$ going to
zero~\cite{Voloshin:1997jh,Lednicky:2003fe}. We focus on 
the function $F$. In this limit, we neglect all components with $1/a_C$
or $1/k^{*}$ in powers greater than 1. We have:
\begin{equation}
F=1+2 \frac{r^{*}} {a_C} + 2 \frac{\vec k^{*} \vec r^{*}} {k^{*} a_C}
+ O\left [\left (\frac {1}{a_C}\right)^2 \right ] + O\left[\left(\frac
{1} {k^{*}}\right)^2\right]+ ... 
\label{eq:Fsmallk}
\end{equation}
which gives the correlation function:
\begin{equation}
C|_{k^*}=\left < A_C F^{*}F \right > \doteq A_C \left ( 1+2 \frac{\left <
r^{*} \right >} {a_C} + 2 \frac {\left < \vec k^{*} \vec r^{*} \right
>} {k^{*} a_C}  
\right ).
\label{eq:CsmallK}
\end{equation}
One notes that:
\begin{equation}
\left < \vec k^{*} \vec r^{*} \right > = \left < k^{*}_{out}
r^{*}_{out} + k^{*}_{L} r^{*}_{L} \cos (\theta^{*} - \psi) \right > =
k^{*} \cos{\psi} \left < r^{*}_{out} \right > .
\label{eq:krav}
\end{equation}
We now consider $C$ at a fixed $k^{*}$ and
$\cos(\psi)$: 
\begin{equation}
C(k^{*}, \cos{\psi}) = A_{C} \left ( 1 + 2 \frac{\left < r^{*} \right
>} {a_C} + 2 \cos{\psi} \frac {\left < r^{*}_{out} \right >} {a_C}
\right ),
\label{eq:Cfixedk}
\end{equation}
and using the uniformity of $\cos{\psi}$ distribution for uncorrelated
particles at small $k^{*}$ we have:
\begin{eqnarray}
C_{+} &\doteq& A_C \int_{0}^{1} C(\cos{\psi}) d\cos{\psi} = 1 + 2 \frac{\left <
r^{*} \right >} {a_C} + \frac {\left < r^{*}_{out} \right >} {a_C}
\nonumber \\
C_{-} &\doteq& A_C \int_{-1}^{0} C(\cos{\psi}) d\cos{\psi} = 1 + 2 \frac{\left <
r^{*} \right >} {a_C} - \frac {\left < r^{*}_{out} \right >} {a_C} \nonumber \\
\frac{C_{+}}{C_{-}} |_{k\to0} &\doteq& 1 + 2 \frac {\left < r^{*}_{out}
\right >} {a_C} . 
\label{eq:drlim}
\end{eqnarray}
These asymptotic values of the double-ratio are shown in
Fig.~\ref{fig:pikdrdep} as arrows. The formula appears to give an easy
way to extract an emission 
asymmetry without the need for tedious analysis. However, one must
remember that in the experiment, the lower the $k^{*}$, the higher is
the experimental uncertainty on the data point. This is because of
statistics, which falls as ${k^{*}}^2$, and because experimental
effects like momentum resolution result in the largest systematic
uncertainty in these bins. Therefore, one has to perform the full
analysis of the double ratio in a broad range of $k^{*}$ to reliably
extract the asymmetry. 

From Eqs.~\eqref{eq:drlim} one concludes that if one restricts the
integrals in the definitions of $C_{+}$($C_{-}$) to a $\cos(\psi)$
range close to $1$($-1$) one will obtain an even larger asymmetry
signal. However, in the experiment, the price to pay is the loss of
statistics, and hence the significance of the signal. It can be shown
that in order to maximize the significance of the signal one should
perform the integrals over the full range of $\cos(\psi)$ (thus
minimizing the statistical error) with the weight
$\cos(\psi)$~\cite{Erazmus:1998xxx,Lednicky:wpcf08}. We 
come back to this crucial point in Section~\ref{sec:cfrep} where we
discuss the spherical harmonics representation of the correlation
function, which happens to naturally introduce similar 
weighting~\cite{Lednicky:wpcf08,Danielewicz:2006hi}.

We finish this chapter by discussing the conventions used in the
analysis. It is important to define and consistently use these
conventions in all steps of the analysis. First, the order of
particles in the pair is important for the definition of $k^{*}$ and
$r^{*}$ because they are defined as the momentum and position of the
{\it first} particle in the pair. We 
adopt a convention that the lighter particle in the pair is always
taken as first. If both particles have equal mass, the positively
charged one is taken as first. The second convention is the definition
of the double ratio, which can either be $C_{+}/C_{-}$, or
$C_{-}/C_{+}$. We chose the former definition. We also note that in
the spherical harmonics representation there is no need for such a
convention. With these definitions
the following general rules hold. Same-charge pairs have correlation
functions going below unity, opposite-charge pairs above unity. That
means that for a given source size asymmetry, the double ratio
for same-sign pairs will be an inverse of the opposite-sign
double-ratio. Finally, with these definitions, a ``double ratio'' below
(above) unity means that the lighter particle is emitted closer to the
center of the system and/or later than the heavier one for
same-sign (opposite-sign) pair.  

\section{Practical implementation of the formalism}
\label{sec:implement}

When using a model that produces individual particles (such as
{\tt THERMINATOR}), the integration in Eq.~\eqref{eq:KonPrat} is
performed via the Monte-Carlo method, iterating over particle
pairs. In that case, the procedure to construct the correlation
function closely resembles the experimental one. It enables the
introduction of some experimental effects, such as acceptance or
momentum resolution, in a straightforward way, as opposed to the
purely analytical models where this is more difficult. Mathematically, the
procedure is a combination of Eqs.~\eqref{eq:cfexp}
and~\eqref{eq:KonPrat}: 
\begin{equation}
C(\vec k^{*}) = \frac {\int A(\vec k^{*}, {\bf r^{*}}) \left | \Psi(\vec
k^{*}, {\bf r^{*}}) \right |^2} { \int B(\vec k^{*}, {\bf r^{*}})} .
\label{eq:cmodweight}
\end{equation}
Note that the $A$ and $B$ now depend on space--time coordinates
${\bf r^{*}}$ as well, because we are using model data in which the
emission points are known. Since we are dealing with pairs of
individual particles we employ the Monte-Carlo procedure, which
replaces the analytic integration by summing over pairs:
\begin{equation}
C(\vec k^{*}) = \frac {\sum_{N pairs} \delta (k^{*}_A - k^{*}) \left |
\Psi(\vec k^{*}, {\bf r^{*}})\right |^2} { \sum_{D pairs} \delta (k^{*}_B - k^{*})},
\label{eq:cmodweightMC}
\end{equation}
where the two sums are performed over two sets of pairs $N$ and $D$ and the
$\delta$ function ensures that only pairs with the correct relative
momentum are taken. Note that there are two possible scenarios, both
of which have slightly different interpretations and uses.
One can perform the calculation using the same pairs for set $N$ and
$D$. Eq.~\eqref{eq:cmodweightMC} is then exactly equivalent
to~\eqref{eq:KonPrat}, but is done via Monte-Carlo
integration. Another option is to take pairs from the same event as 
sample $N$ and pairs of particles from different events as sample
$D$. In that case, 
the correlation function $C$ contains not only the femtoscopic effect
but also all other event-wide correlations which are present in the 
model, projected to the two-particle space. It is therefore very
useful for the experimentalists who can study the differences between
the two to estimate the magnitude of the non-femtoscopic effects. The two
cases have one more important difference. Usually the CF is
constructed as a ratio of two histograms: $N$, a signal, which is
filled with the weight $\left | \Psi \right |^2$ for every pair, and $D$, the background,
which is filled with unity for each pair. The errorbars on $C=N/D$
have different meaning in the two cases. In the first one it is just
the spread of the weight in a given sample. In the second case it is a
true errorbar, comparable to the experimental one, since samples $N$
and $D$ are statistically independent. In this work we are using the
first way to calculate the correlation function, since we are not
interested in non-femtoscopic correlations, but rather in physics
effects accessible via femtoscopy.

\subsection{Correlation function representation}
\label{sec:cfrep}

Femtoscopic correlation functions have been represented in two main
forms: as a 1-dimensional histogram with the magnitude of the relative
momentum $k^{*}$ or $q=2k^{*}$ on the axis, or as a 3-dimensional
histograms with $k^{*}_{out}$, $k^{*}_{side}$ and $k^{*}_{long}$ on the
axes. For identical pion correlations it is also useful to use a 3D
histogram with the relative momentum components $q_{out}$, $q_{side}$
and $q_{long}$ calculated in the LCMS. Note that out, side and long
decomposition is possible also for pairs of non-identical particles;
one has to use the generalized four-momentum variable $\tilde q$
instead of $q$: $\tilde q=q-P(qP)/P^2$. The first form requires minimal
statistics, but only allows to determine the 1D overall source
size. The second one allows for the determination of sizes in all 3
directions, but requires significant statistics. Up to now, the
non-identical correlation function was represented as a set of two
1-dimensional histograms - one for $C_{+}$ and one for $C_{-}$ with
respect to the ``out'' direction. This allowed for the determination
of 1D source size and a study of the ``double ratio'' to access
asymmetries. However, if one wanted to study ``double ratios'' for
other directions (``side'' and ``long''), one needed to create
separate sets of correlation functions. 

Recently, a more advanced way of representing the correlation function,
the Spherical Harmonics (SH), was proposed~\cite{Chajecki:2008vg}. It
has several important advantages. It encodes the full 3D information of
the correlation in a set of 1D plots. Generally, this does not need
to be an advantage, because a perfect representation of all the
features of any 3D function requires an infinite set of $l,m$ components
(meaning: infinite set of 1D histograms). But it so happens that the
intrinsic symmetries of a pair distribution in the femtoscopic  
analysis result in most of the $l,m$ components to vanish. It has
also been shown that out of these that do not vanish, only those with
small $l$ contain important information, which means that one can
safely truncate the decomposition at a rather small $l$ without the
fear of losing any physics information. It is as if the spherical
harmonics have been specifically designed to efficiently represent a
femtoscopic correlation function.

The spherical harmonics representation has the advantages of both the 
1D correlation function (because it requires less statistics) and a 3D
one (because it encodes the important part of the 3D
information). However, first attempts to apply the decomposition
methods, which were developed for identical particles, to the
non-identical particle correlations were not 
successful. Essentially, one had to first construct the numerator and
denominator as 3D histograms (usually in $|\vec q|$, $\cos \theta_q$
and $\phi$), then divide them and decompose the resulting 3D
correlation function, negating the low-statistics advantage of the 1D
representation. The decomposition procedure also relied heavily 
on the symmetries present for identical correlations. This presented a
problem for non-identical correlations, because their primary goal is to
study the emission asymmetry. This breaks one of the symmetries
present in identical analysis. Also, the single particle acceptance of
some experiments produced ``acceptance holes'' i.e. regions of empty
bins in the 3D function. Also, statistics for 
non-identical pairs were significantly lower than for pion-pion
correlations, so filling all the bins in a 3D 
correlation function with significant number of pairs became a
challenge. To solve these problems, a new technique to represent the
correlation function in spherical harmonics was
developed~\cite{Kisiel:2009iw}. Both the numerator and denominator are
stored directly in spherical harmonics ({\it not} as 3D histograms),
and the procedure to calculate the CF directly from them (again, {\it
not} involving any 3D histogram) has been
presented~\cite{Kisiel:2009iw}. An additional benefit of the method is
that the covariances between all $l,m$ 
components are explicitly taken into account. Unless otherwise noted,
all further correlation functions presented in this work are
represented using this method. We also recommend that experimental
groups use this method.  

As was the case for identical particle femtoscopy, the spherical
harmonics representation turned out to have specific synergies with
the non-identical femtoscopic correlation analysis. The important
femtoscopic information is contained in only two $l,m$ components,
while the other two can be used as additional cross-checks of experimental
procedures. If one wishes to analyze the full 3D information, one needs
to analyze only two more $l,m$ components -- the rest can be
essentially neglected as they should be consistent with zero, or they
will not contain additional useful information. 

We now investigate the important components of the SH decomposition
and their sensitivity to femtoscopic information. We perform a
simple calculation in which we assume the source to be a 3D Gaussian
in the LCMS, having three different sizes $R$ in three directions
($out$, $side$ and $long$). It also has a non-zero mean value
$\mu_{out}$ in the $out$ direction: 
\begin{equation}
S({\bf r}) = \exp \left ( - \frac {(r_{out} - \mu_{out})^2}
{R_{out}^{2}} - \frac {{r_{side}}^2} {R_{side}^{2}} - \frac
{{r_{long}}^2} {R_{long}^{2}} \right ).
\label{eq:stest}
\end{equation}
With this source function we perform the
integration~\eqref{eq:KonPrat} to calculate the correlation function 
and present it in spherical harmonics representation. We expect that
the main femtoscopic information is contained in the following
components: $C_{0}^0$, $\Re{C_{1}^{1}}$, $\Re{C_{0}^{2}}$ and
$\Re{C_{2}^{2}}$.  

\begin{figure}[tb]
\begin{center}
\includegraphics[angle=0,width=0.48 \textwidth]{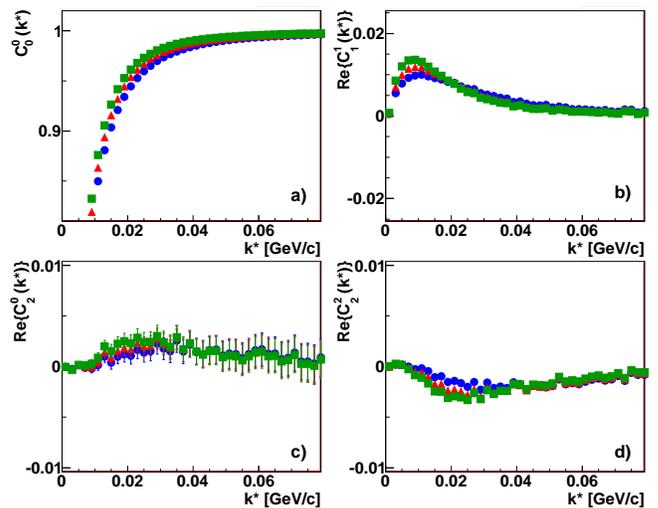}
\end{center}
\vspace{-6.5mm}
\caption{(Color on-line) SH components of the correlation function for
$R_{out} = 8$~fm, $R_{side}=R_{long}=4$~fm, $\mu_{out} = -4$~fm (blue
dots), $R_{out} = 10$~fm, $R_{side}=R_{long}=5$~fm, $\mu_{out} = -5
fm$ (red triangles), $R_{out} = 12$~fm, $R_{side}=R_{long}=6$~fm,
$\mu_{out} = -6$~fm (green squares).
\label{fig:shdeprad}}
\end{figure}

\begin{figure}[tb]
\begin{center}
\includegraphics[angle=0,width=0.48 \textwidth]{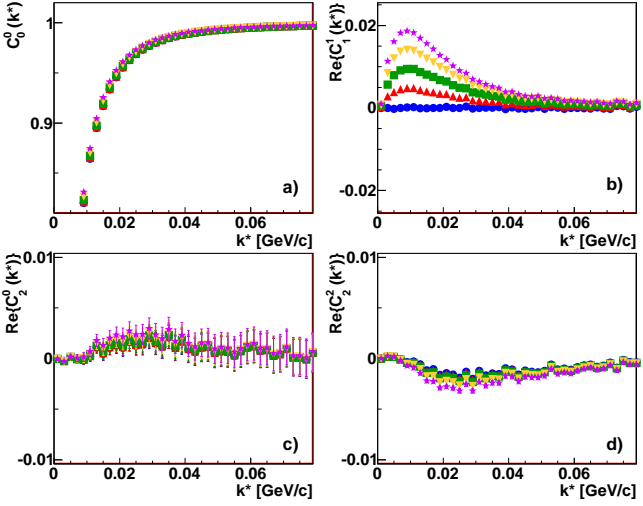}
\end{center}
\vspace{-6.5mm}
\caption{(Color on-line) SH components of the correlation function for
$R_{out} = 10$~fm, $R_{side}=R_{long}=6$~fm. Emission asymmetry
$\mu_{out}$ is changed from $0$~fm (blue dots) via $-2$~fm (red
up-triangles), $-4$~fm (green squares), $-6$~fm (yellow
down-triangles), up to $-8$~fm (violet stars).  
\label{fig:shdepshift}}
\end{figure}

To study the sensitivity to the source size we increase the overall
size, but keep the radii ratios the same. The results are plotted in
Fig.~\ref{fig:shdeprad}. All components show some sensitivity, but the
$C_0^0$ is affected the most.  
In the next step (shown in Fig.~\ref{fig:shdepshift}) we keep the
source size the same, while we increase the emission asymmetry
($\mu_{out}$). The main sensitivity is in the $\Re C_1^1$
component. More importantly, no asymmetry results in vanishing $\Re
C_1^1$, while increasing the asymmetry increases the signal in $\Re C_1^1$
monotonically, approximately linearly. $\Re C_1^1$ is functionally
equivalent to the ``out'' double-ratio. Obtaining a quantitative as
well as qualitative information about the asymmetry should be possible
from the analysis of it (in correlation with at least $C_0^0$ 
where sensitivity is also seen, but to a smaller degree). In addition,
from Eq.~\eqref{eq:drlim}, we concluded that in order to maximize the
significance of the asymmetry signal one should integrate the
correlation function with the weight equal to
$\cos(\psi)$~\cite{Erazmus:1998xxx}. Remarkably, 
the definition of $\Re C_1^1$ is essentially:
\begin{equation}
\Re C_1^1(q) = N \int C(q, \cos(\theta_q), \phi_q) \sin(\theta_q)
\cos(\phi_q) d\Omega_q ,
\label{eq:c11def}
\end{equation}
where $N$ is the normalization factor, while $\theta_q$ and $\phi_q$ are
the longitudinal and transverse components of the $\psi$ angle.
The $\Re C_1^1$ component happens to be the optimal way to maximize
the {\it transverse} components of the asymmetry signal, due to its
$\cos(\phi_q)$ weight, while the $\Re C_1^0$ maximizes the
longitudinal asymmetry signal, due to its $\cos(\theta_q)$ weight:
\begin{equation}
\Re C_1^0(q) = N \int C(q, \cos(\theta_q), \phi_q) \cos(\theta_q)
d\Omega_q .
\label{eq:c11def}
\end{equation}
Once again it appears as if the spherical harmonics were specifically
designed for the femtoscopic correlation function representation.

\begin{figure}[tb]
\begin{center}
\includegraphics[angle=0,width=0.48 \textwidth]{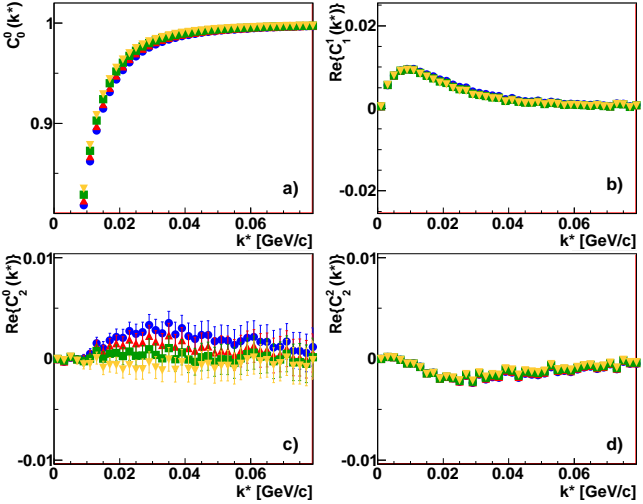}
\end{center}
\vspace{-6.5mm}
\caption{(Color on-line) SH components of the correlation function for
$R_{out} = 10$~fm, $R_{side}=4$~fm, $\mu_{out} = -4$~fm. $R_{long}$ is
changed from $4$~fm (blue dots) via $6$~fm (red up-triangles), $8$~fm
(green squares) to $10$~fm (yellow down-triangles).
\label{fig:shdeplong}}
\end{figure}

Next, in Fig.~\ref{fig:shdeplong} we keep the transverse source size
and the asymmetry the same, while we change the $R_{long}$
radius. Obviously we see a change in the $C_0^0$ which reflects the
growth of the overall system size. But the most sensitive component is
$\Re C_2^0$ which carries information about the ratio of the
transverse to the longitudinal radii, due to its 
$\cos^2(\theta_{q})$ weighting.

\begin{figure}[tb]
\begin{center}
\includegraphics[angle=0,width=0.48 \textwidth]{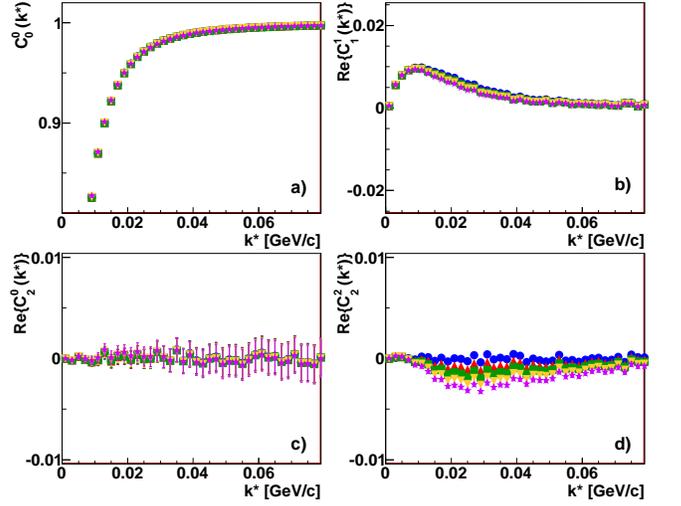}
\end{center}
\vspace{-6.5mm}
\caption{(Color on-line) SH components of the correlation function for
$R_{long} = 8$~fm, $\mu_{out} = -4$~fm. Transverse radii sum is kept
constant: $R_{out} = 7$~fm, $R_{side}=8$~fm (blue dots); $R_{out} =
8$~fm, $R_{side} = 7$~fm (red up-triangles); $R_{out} = 9$~fm,
$R_{side} = 6$~fm (green squares); $R_{out} = 10$~fm, $R_{side} =
5$~fm (yellow down-triangles); $R_{out} = 11$~fm, $R_{side} = 4$~fm
(violet stars).  
\label{fig:shdepperp}}
\end{figure}

Finally (see Fig.~\ref{fig:shdepperp}) we keep the sum of the
transverse radii the same, but we change their ratio. The $\Re C_2^2$
component, with its $\cos^2(\phi_{q})$ weighting, is the most
sensitive to these changes, while the others remain practically
constant.  

The above calculations show that by analyzing just two components of
the SH decomposition ($C_0^0$ and $\Re C_1^1$) one can already perform a
meaningful femtoscopic analysis and determine the overall source size
and emission asymmetry. Adding just two more components ($\Re C_2^0$ and
$\Re C_2^2$), one can also determine source radii in all 3 directions,
which would normally require a full analysis of a 3D correlation
function. Full 3D analysis does require larger statistics than the 1D
size+asymmetry one, as the sensitivity to the observables in the $l=2$
components is smaller than in the $l=0$ and $l=1$ ones.
One should stress that in the SH representation one is dealing
with 200 data points (50 points per histogram, 4 components) versus the
625000 bins in the 3D histogram representation. The huge savings in
computation time and method complexity do not compromise the physics -
one is able to obtain essentially the same femtoscopic information (3
sizes and emission asymmetry). It means that the 3D representation
is a particularly inefficient way of storing the femtoscopic
information, while the spherical harmonics one seems to be perfectly
tailored for that task. We add that in all theoretical
calculations that we have done all the other components were either
required to vanish from symmetry relations, were consistent with zero,
or contained femtoscopic information that was already accessible via
the four main components. In
experiment one should, in addition, look at the $\Re C_1^0$
(equivalent to the ``long'' double ratio) and $\Im C_1^1$ (equivalent
to the ``side'' double-ratio). The former is expected to show zero
asymmetry, the latter is required to vanish due to symmetry reasons,
but their deviations from zero may signal experimental reconstruction
problems.

\subsection{Extracting qualitative information}
\label{sec:fitting}

In the previous paragraph the sensitivity of the non-identical
particle correlation function to the source size and asymmetry was
illustrated. Dependencies in Figs.~\ref{fig:shdeprad},
\ref{fig:shdepshift}, \ref{fig:shdeplong}, \ref{fig:shdepperp} 
show that source size parameters influence all SH components at
the same time, and it is not possible to analyze them separately. In
identical particle femtoscopy obtaining Gaussian source size parameters
is straightforward: the integral in Eq.~\eqref{eq:KonPrat} can be
performed analytically if one assumes that $S({\bf r})$ is a 3D
Gaussian and does not depend on pair momentum $\vec K$ (although the
pairs transverse momentum $K_{T}$ dependence can be recovered by
$K_{T}$ binning; the rapidity binning is also possible). To take into
account the Coulomb interaction, an 
approximate wave-function is used in which the Coulomb part is
factorized out and replaced by an averaged function that only depends
on the magnitude of $k^{*}$. This procedure~\cite{Sinyukov:1998fc},
enables to write a simple analytic formula, which can be directly
fitted to the 3D correlation function and provides femtoscopic radii.

\begin{figure}[tb]
\begin{center}
\includegraphics[angle=0,width=0.4 \textwidth]{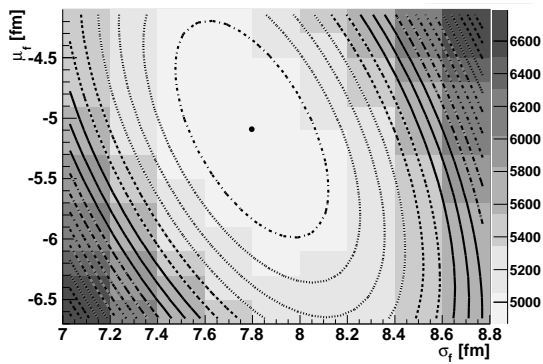}
\end{center}
\vspace{-6.5mm}
\caption{(Color on-line) An example of the 2D $\chi^2$ map obtained by
fitting a non-identical particle correlation function. Underlying
colored histogram represents the ``mesh'' obtained in the fitting
procedure. The lines represent a fitted 2D parabola. The point
represents the minimum of the parabola - the best fit value.
\label{fig:chi2mapexample}}
\end{figure}

In non-identical particle femtoscopy the Coulomb interaction is the
source of the correlation, so it cannot be factorized out. Performing
the integral in Eq.~\eqref{eq:KonPrat} analytically
becomes impossible, especially when one needs to consider the strong
interaction as well. The procedure must be carried out
numerically. One starts with the same assumption as for identical
particle femtoscopy. The source is a 3D
Gaussian~\eqref{eq:stest}, where the additional modification allowing
for a non-zero shift in the ``out'' direction is introduced. One then
assumes a certain set of source parameters $(R_{out}, R_{side},
R_{long}, \mu_{out})$ and calculates the corresponding correlation
function, according to Eq.~\eqref{eq:KonPrat} and using the corresponding
wave function~\eqref{eq:fullpsi}. As the source function is assumed to be
momentum independent, one needs an input momentum distribution (the
assumption of momentum independence of the source function is not
required provided the selected $(p_{T},y)$ is sufficiently narrow). This
can be achieved by e.g.\ taking particles' momenta from real
pairs from data and assigning the separation ${\bf r^{*}}$ randomly
generated from~\eqref{eq:stest}. The resulting ``model'' CF is
compared to the ``measured'' one via a $\chi^2$ test. The procedure is
repeated for several sets of source parameters to find the one that
fits the ``data'' best. This set is taken as the result of the
fit. The  procedure can be refined by ensuring that the ``model''
calculation is done for points which form a regular ``mesh'' in
parameters' space. In the simplest one parameter fit, one obtains the fit
value and error from the location of $\chi^2$ minimum on the
parabola. When 2 independent parameters are fitted (the most common
case for non-identical correlations, when we usually fit overall radius $R$
and emission asymmetry $\mu$) one creates a 2D ``mesh'', to which one
can fit a 2D parabola - obtaining the best fit parameters, their
errorbars as well as the covariance between them. An example of such
analysis is shown in Fig.~\ref{fig:chi2mapexample}. One can also
employ a minimization package (e.g.\ Minuit) to perform the fitting
process. The number of points in the CF ``mesh'' can be kept at
minimum by using a simple linear or quadratic interpolation between
points~\cite{Lednicky:2005tb,Lednicky:1999nn}. 

For the results presented in this work we used a software package
{\tt CorrFit} to perform the numerical fitting procedure described
above. It was developed for the analysis of non-identical particle
correlations in the STAR experiment~\cite{Kisiel:2004cf,Adams:2003qa}. 
It allows for significant flexibility when choosing various
parameters of the fitting process. For this work we have made the
following choices: (1) The input data were correlation functions in
spherical harmonics representation. Only $C_{0}^{0}$ and $\Re
C_{1}^{1}$ components were fitted; the covariance between the two was
taken into account. (2) The particular form of the $S$ from
Eq.~\eqref{eq:stest} was taken with the additional constraints that
$R_{side}$ was assumed to be equal to $R_{out}$, while $R_{long} = 1.3
R_{out}$, following the identical particle 3D femtoscopic results for
pions from RHIC~\cite{Adams:2004yc}. Therefore, only two independent
parameters were fitted: $R_{out}$ and $\mu_{out}$. (3) The input
momentum distributions were taken from the STAR experiment, ensuring
that the momentum acceptance was the same. (4) The standard package
from R.~Lednicky was used to calculate the pair
weights~\cite{Lednicky:2005tb,0954-3899-35-12-125109}. The input 
theoretical correlation functions were calculated taking into account
Coulomb interaction only. In such case only the Coulomb part of the
pair weight was calculated in the fitting procedure. (5) One had the
possibility to introduce ``pair purity'' and momentum resolution
corrections. Details of the purity correction will be discussed later
in this Section. The momentum resolution correction was not necessary
for model studies shown in this paper.

\subsection{Influence of pair purity}
\label{sec:purity}

For femtoscopic analysis one of the main experimental issues is the
``purity'' $P$ of the analyzed pair sample, that is the fraction of pairs
in the sample that should be treated as ``femtoscopically
correlated''. There are several reasons why a pair of particles should
not be treated as correlated. From the experimental side, it may
happen that one (or both) of the particles in the pair has been
misidentified (this is why $P$ is traditionally called
``purity''). Another common scenario is when one of the particles is a
product of a weak decay. In that case, it is the ``parent'' particle
that should be treated as ``femtoscopically correlated'', leading in
some cases to rather complicated cases of ``residual correlations'',
that is feed-down of femtoscopic correlation between parent particles
into the daughters' correlation function. Recently, such cases have
been studied in detail in baryon--baryon
correlations~\cite{Adams:2005ws,Gos:2007cj}. However, in systems
considered in this paper, containing a meson, residual correlations
are smeared away by the decay momenta, so such pairs should be treated
as ``not femtoscopically correlated''. These experimental components
of $P$ are clearly experiment-dependent and it is up to the experiment
to correct for such effects (or at least provide a realistic estimate
of $P$). We will not address it further in this work.

However, another contribution to $P$ remains, the estimate of which
is model dependent. If one of the particles in the pair comes from
a strongly decaying resonances that lives very long, e.g.\ the $\omega$
meson, we may need to treat it as ``not femtoscopically
correlated''. That is because the source of the correlation, the pair
wave function $\Psi$ is usually peaked at low $r^{*}$ values, while at
large values it produces no correlation. This can be dealt with in two
ways. First is to assume, in the data analysis, a source function that
perfectly describes such long-range $r^{*}$ tails. However, this is
difficult to do, and it is very dependent on the model that we use to
model such tails. Another solution is to treat the particles in the
tails as ``not femtoscopically correlated'' -- in other words to
provide a model estimate of the decrease of $P$ coming from long-lived
resonances. This is usually done by assuming that a source is a 3D
sphere with a Gaussian profile, and counting any pair which is
outside this sphere as non-correlated. We note that the
``non-gaussian effects'' have been seen in all RHIC
experiments in the identical particle
correlations~\cite{Adams:2004yc,Adler:2004rq}. They have been
attributed, at least in part, to long-lived
resonances~\cite{Kisiel:2006is}, and recently, new techniques have been
proposed to analyze them in detail~\cite{Adler:2006as}. Such analysis
are not yet possible in non-identical particle analysis, so we limit
ourselves to the simple model estimation of $P$. 

We characterize the overall ``purity'' of a pair sample by a
percentage of pure pairs $P$. It can be (and in the experiment it 
usually is) a function of $\vec k^{*}$. Assuming that the ``non-pure''
pairs are not correlated, the correlation function can be trivially
corrected for purity:
\begin{equation}
C_{corrected}(\vec k^{*}) = (C_{measured}(\vec k^{*})-1)/P(\vec k^{*})
+ 1.
\label{eq:purcor}
\end{equation}
Employing the formula~\eqref{eq:purcor} requires a precise knowledge of
the fraction of correlated pairs $P$ in the measured sample. Any
uncertainty in its value will be a source of the systematic error. In
contrast to the identical particle femtoscopy, $P$ cannot be easily
inferred from the fit, independently of the source parameters $\sigma$
and $\mu$. By inspecting Fig.~\ref{fig:pikcfdep} and
Eq.~\eqref{eq:purcor} one concludes that
lowering purity influences the correlation function in a way that is
similar to the changes introduced by varying the source size. One can
obtain a satisfactory fit to the correlation function by adjusting
either one of these parameters, so treating them both as free makes
them highly correlated. While possible, the independent determination
of $\sigma$ and $P$ requires significant statistics and very good
control of the non-femtoscopic
background~\cite{Lednicky:2005tb,Lednicky:1999nn}.  

\begin{figure}[tb]
\begin{center}
\includegraphics[angle=0,width=0.48 \textwidth]{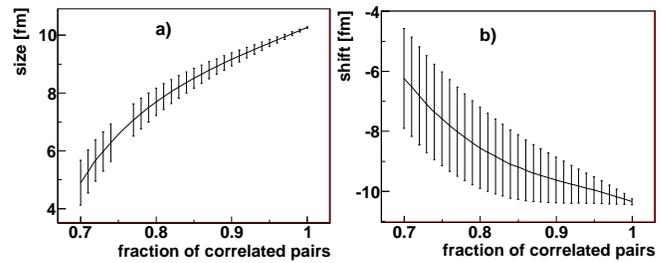}
\end{center}
\vspace{-6.5mm}
\caption{Values of the correlation function fit (size a) and shift b)
  ), as a function of a pre-defined fraction of primary pairs.
\label{fig:purdep}}
\end{figure}

Usually, the practical solution is to come up with the best estimate of
$P$ by other means (both experimental and theoretical), and fix its
value for the fitting. By performing the procedure for several fixed
values of $P$, within the reasonable uncertainty range, one obtains
the systematic error on the fit values coming from the purity
estimate. As an example on Fig.~\ref{fig:purdep} we show how the fit
values for the same input correlation function change as one adjusts
the fixed pair purity $P$. Within the reasonable range of $P$
variations of the order of 20\% the dependence of $\sigma$ and $\mu$
on $P$ is monotonic and noticeable. We expect that the purity estimate
will be the major source of the systematic error. We will address the
theoretical part of the estimates of pair purity in
Section~\ref{sec:charmodelinput}.  

\subsection{Influence of the momentum resolution}

Momentum resolution will influence the femtoscopic correlation
function in a well defined way. Since reconstructed momenta will
differ from the true ones, the correlation effect (visible as either a
``peak'' or a ``depression'' at low $k^{*}$)  will be reduced. The
Coulomb correlation is rather sharp at low relative momentum, so one
expects the small $k^{*}$ points to be influenced the most. The
``double-ratio'' will be affected as well. One can study the influence
of the momentum resolution by performing a theoretical calculation in
the following way. One calculates $C$ according
to~\eqref{eq:cmodweightMC}, but calculates two sets of relative momenta
$k^{*}$ for each pair. The first one $k^{*}_{true}$ is used to
calculate $\Psi$, the second one $k^{*}_{smeared}$ is used to
determine the correlation function bin in which the pair is
stored. The second set is calculated by smearing the particles'
momenta with the parametrization of the momentum resolution obtained
from the experiment. By comparing the ``unsmeared'' and ``smeared''
correlation functions one can judge the importance of the momentum
resolution effect. One can also employ the same technique when
calculating the ``theoretical'' correlation functions during the
fitting procedure described in Section~\ref{sec:fitting}. In this way
the fitting procedure automatically corrects for momentum
resolution. The precise determination of the momentum resolution
parameters is very dependent on the experimental features. We will not
discuss it further in this work (see, e.g.~\cite{Lednicky:1999nn}).

\section{Systematic checks of the method}
\label{sec:crosschecks}

In the previous sections we have made qualitative claims that the
analysis of the non-identical particle correlations should enable the
estimation of the source size and emission asymmetry for pions, kaons
and protons emitted in the heavy-ion collision. In this section we aim
to show the quantitative checks of the method. We focus on answering
two questions: Can the method reliably and quantitatively recover the
source size and emission asymmetry? If yes, what systematic
uncertainties in this estimation come from the method itself?

We emphasize that in this Section we focus only on the technical
aspect of the non-identical particle correlations method. So, the
{\tt THERMINATOR} calculations are used only as ``test samples''. 

\subsection{Details of the procedure}

We chose the following procedure: we simulate heavy-ion Au+Au
collisions with the {\tt THERMINATOR} model. We do it for 6 centrality bins:
0-5\%, 5-10\%, 10-20\%, 20-30\%, 30-40\% and 40-50\%. We calculate the
full set of non-identical correlation functions: same and opposite
sign pion--kaon, same and opposite-sign pion--proton and same sign
kaon--proton for these centralities, producing 6x5=30 independent
correlation functions. Then, we proceed to treat these functions as if
they were experimental ones: applying the purity correction and
fitting them with the {\tt CorrFit} software. In the end we compare the
input values from the model with the fit results and see if they
match.

We simulated 50K events for each centrality. The parameters of the
model were the same as the ones used
in~\cite{Kisiel:2008ws,Broniowski:2008vp}, which are known to reproduce
both single particle spectra and overall femtoscopic sizes from
identical particle correlations. The correlation functions were
calculated by combining particles from these events into pairs and
employing Eq.~\eqref{eq:cmodweightMC}. Spherical harmonics representation
of the CF was used. Then, each function was fitted with the {\tt CorrFit}
program, assuming that the two-particle source is a 3D Gaussian in
LCMS, according to Eq.~\eqref{eq:stest}. The sideward and longitudinal
sizes of the source were fixed to be equal to $R_{out}$ and
$1.3 R_{out}$ respectively, according to the results obtained for 
identical pions. In this way the fitting procedure had only two
parameters: overall size $R_{out}$ and emission asymmetry
$\mu_{out}$. We note that while the full 3D analysis, with $R_{side}$
and $R_{long}$ as independent free parameters of the fit, is in
principle possible, we do not discuss it in this paper, as we consider
it less interesting: it would provide 3D information about the source,
but with much lower precision than the 3D identical pion
analysis. Instead, we chose to focus on asymmetry, which is a unique
observable accessible only via non-identical particle femtoscopy. 

\subsection{Characterizing model input}
\label{sec:charmodelinput}

\begin{figure}[tb]
\begin{center}
\includegraphics[angle=0,width=0.48 \textwidth]{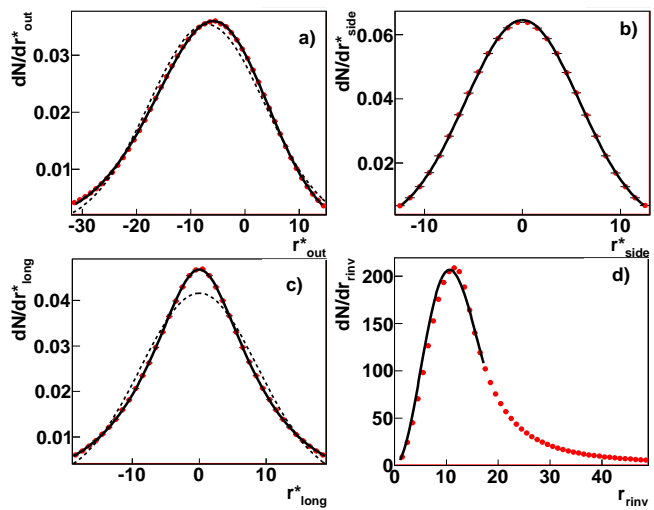}
\end{center}
\vspace{-6.5mm}
\caption{Two-particle emission function for pion--kaon pairs at
centrality 10-20\%. $r^{*}_{out}$ in a), $r^{*}_{side}$ in b),
$r^{*}_{long}$ in c), $r_{inv}$ in d). Lines are fits to the
distributions, see text for details.  
\label{fig:exampleS}}
\end{figure}

Let us inspect a typical two-particle emission function $S_{AB}$
coming from the {\tt THERMINATOR} model, shown in
Fig.~\ref{fig:exampleS}. The distributions in $out$, $side$ and $long$
directions show different behavior. Only the $side$ one is well
described by a Gaussian; attempts to fit the $out$ and $long$ by
Gaussians are shown as dashed lines. The $long$ can be described by an
exponential hyperbola function:
\begin{equation}
f_{EH} = \exp \left (-\sqrt{ \frac {{r^{*}_{long}}^{2}} {\sigma_{long}^2} +
\alpha^2 }\right ),
\label{eq:feh}
\end{equation}
where $\sigma$ and $\alpha$ are parameters. For the description of the
$out$ one needs an asymmetric exponential hyperbola:
\begin{equation}
f_{EH}^{as} = \exp \left ( -\sqrt{ \frac {(r^{*}_{out}-\mu_{out})^{2}}
{\sigma_{out}^2} + \alpha^2 } \right )(1 + \zeta (r^{*}_{out} -
\mu_{out})) ,
\label{eq:fehas}
\end{equation}
where $\sigma_{out}$, $\mu_{out}$, $\alpha_{out}$ and $\zeta$ are
parameters. We have confirmed that these functional forms are general,
that is, one 
is able to fit them to the emission functions at all centralities and
for all pair types. In contrast, the fitting procedure that we have
just described in the previous Section assumes a proper Gaussian
source distribution. Therefore, one needs 
to find a suitable variable which will enable the comparison between
the model input, which is non-Gaussian, with the fit output, which is
postulated to be a Gaussian. In principle, one can think of removing
the Gaussian assumption and actually using the functional forms
mentioned above for the fitting. This is certainly possible, both from
the point of view of the technique as well as the software
tools. However, it brings 
two potential problems. First of all, while the functional forms are
able to fit the {\tt THERMINATOR} data very well, there is no reason to
assume it will be the case for other models. So, by choosing these
particular forms one would introduce a strong model
dependence. Moreover, it is not known how the parameters of these
functions could be compared to the Gaussian radii from identical
particle femtoscopy. In this work we access the overall
directionally-averaged size of the system only, so the details of the
source  function dependence in the three directions separately will be
lost. In this work we decided to use the Gaussian assumption, having
in mind that it may introduce systematic effects, when  comparing the
fit values to input two-particle distributions. 

We are then presented with the following situation: the
``experimental'' fit procedure produces the direction-averaged source
size $R$ and the emission asymmetry in $out$ direction $\mu_{out}$. On
the other hand, to fully describe the theoretical model input, one
needs the following parameters: $\sigma_{out}$, $\alpha_{out}$,
$\zeta$, $\mu_{out}$, $\sigma_{side}$, $\sigma_{long}$ and
$\alpha_{long}$. To compare the two, one needs to find common variables
which can be compared, and which will tell whether our ``experimental
fit'' reproduces the ``model input''.

\begin{figure}[tb]
\begin{center}
\includegraphics[angle=0,width=0.4 \textwidth]{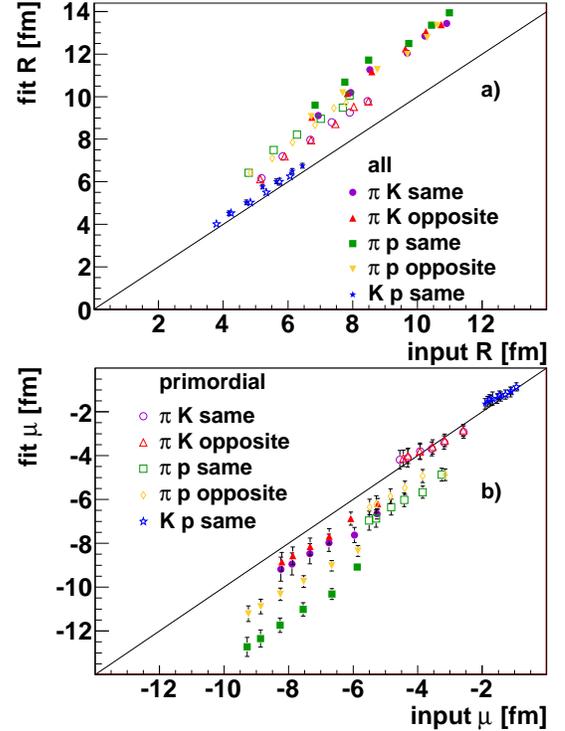}
\end{center}
\vspace{-6.5mm}
\caption{(Color on-line) The correlation between input values obtained
from model and fit values. Panel a) shows overall system size, b) -
emission asymmetry in the $out$ direction. Open points show
calculations for primary particles only, closed points - for all 
particles. Violet circles are same-sign pion--kaon, red up-triangle:
opposite-sign pion--kaon, green squares: same-sign pion--proton, yellow
down-triangles: opposite-sign pion--proton, blue stars: same-sign
kaon--proton. 
\label{fig:inpfitcor}}
\end{figure}

\begin{figure}[tb]
\begin{center}
\includegraphics[angle=0,width=0.48 \textwidth]{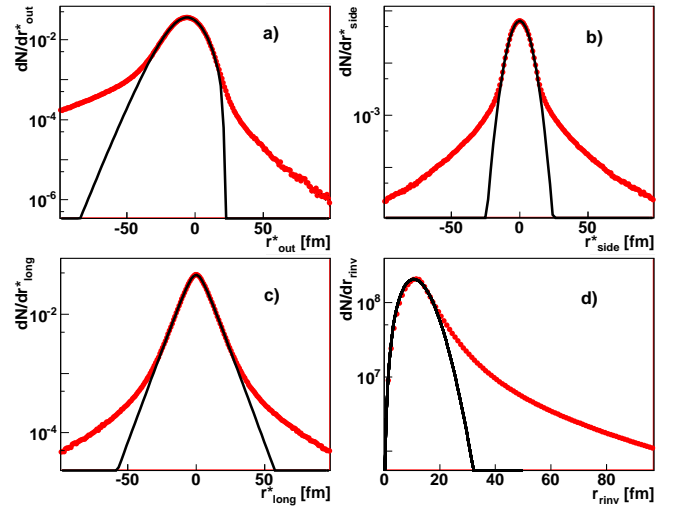}
\end{center}
\vspace{-6.5mm}
\caption{An illustration of the long-range tails produced by
{\tt THERMINATOR}. Points: separation distributions for pion--kaon pairs,
a): $r_{out}$, b): $r_{side}$, c): $r_{long}$, d) $r_{inv}$. Lines are
fits to the region near the peak, using the functional forms:
$f_{EH}^{as}$ for $out$, Gaussian for $side$, $f_{EH}$ for $long$ and
$f_{G}$ for $inv$.  
\label{fig:longtail}}
\end{figure}

In traditional HBT one compares 1-dimensional femtoscopic radii,
defined as a $\sigma$ of a 1D-Gaussian approximation of the
single-particle emission function (usually in PRF). We wish to define
an analogous variable for non-identical study, but defined in LCMS to
facilitate the comparisons to the 3D identical particle interferometry
results. The task is complicated by  the fact, that neither the
``model input'' nor the ``experimental fit'' produce a source which is
a perfect Gaussian. Nevertheless, one can plot the source distribution
$dN/dr$ in both cases. Then, one fits the distribution with the
Gaussian formula, multiplied by the proper Jacobian: 
\begin{equation}
F_G(r) \approx \exp(-\frac {{r}^2} {2 R_{av}^{L}}) {r}^2.
\label{eq:gausjacob}
\end{equation}
As will be discussed later, one must restrict the range of this fit to
the low-$r$ values, as these are the ones that contribute to the
femtoscopic effect. The contributions from the large-$r$ part of
the source must be dealt with separately. The lower--right part of
Fig.~\ref{fig:exampleS} shows the example of the model $r_{inv}$ 
distribution; the line is a fit according to~\eqref{eq:gausjacob},
done in the range $0-20$~fm. Since both $r_{out}$ and $r_{long}$
distributions are manifestly non-Gaussian, so is the $r_{inv}$
distribution\footnote{It can be shown that $r_{inv}$ distribution is a
Gaussian only if all three $r_{out}$, $r_{side}$ and $r_{long}$ are
Gaussians and additionally all three have the same width.}. Therefore,
the fitted $R_{av}^{input}$ is only characterizing the general size of
the system. The same fitting procedure can be applied to the $r_{inv}$
distribution produced by the ``experimental fit'', producing the
$R_{av}^{fit}$ ``experimental'' value. For a more detailed discussion
of the approximate relations between 1D and 3D source size parameters,
the relations between values in PRF and LCMS as well as a more
detailed description of estimating $R_{av}$ we refer the reader to
Appendix~\ref{sec:emfuncs}.  

One also needs to define, what does ``emission asymmetry'' $\mu_{out}$
mean in the case of the model. One can come up with several definitions:
(a) the ``mean'' of the $r^{*}_{out}$ distribution, (b) the ``mean''
of the $r^{*}_{out}$ distribution, but only taking into account parts
of the distribution near to the peak (neglecting the long-range tail),
(c) the position of the maximum in the $r^{*}_{out}$ distribution. All
three are correlated, but (a) shows large sensitivity to long-range
tails and therefore is not well defined. Out of
the other two (b) should be exactly the same as $\left < r^{*}_{out}
\right >$, which is the value used in the theoretical formalism, while
the relation of (c) to $\left < r^{*}_{out} \right >$ is less
straightforward. So, we chose to use the definition (b).

\begin{table}[tb]
\caption{Estimate of the theoretical purity $P$ for all pair types,
versus centrality. See text for details.
\label{tab:theorypur}}
\begin{tabular}{|r|rrrrrr|}
\hline
$c$ [\%] & 0-5  & 5-10 & 10-20 & 20-30 & 30-40 & 40-50 \\
\hline
$\pi K$ & 0.83 & 0.82 & 0.80 & 0.76 & 0.73 & 0.70   \\
$\pi p$ & 0.80 & 0.79 & 0.77 & 0.75 & 0.73 & 0.71   \\
$K p$   & 0.96 & 0.95 & 0.95 & 0.94 & 0.93 & 0.93   \\
\hline
\end{tabular}
\end{table}

In Fig.~\ref{fig:inpfitcor} the comparison between ``input'' and
``fitted'' radii is shown, for all considered pairs,  both for
primordial and all particles. A clear correlation is seen between the
two. However, most of the results do not lie on the ``perfect'' $x=y$
line. Within each pair system the correlation is clear and
monotonic. Moreover, the deviations from the ``perfect'' curve seem to
be a systematic shift, similar for all points of the same pair
type. The shift from perfect values for a given pair type is seen both
for the $R_{inv}$ and the $\mu_{out}$ variables. 

\begin{figure}[tb]
\begin{center}
\includegraphics[angle=0,width=0.4 \textwidth]{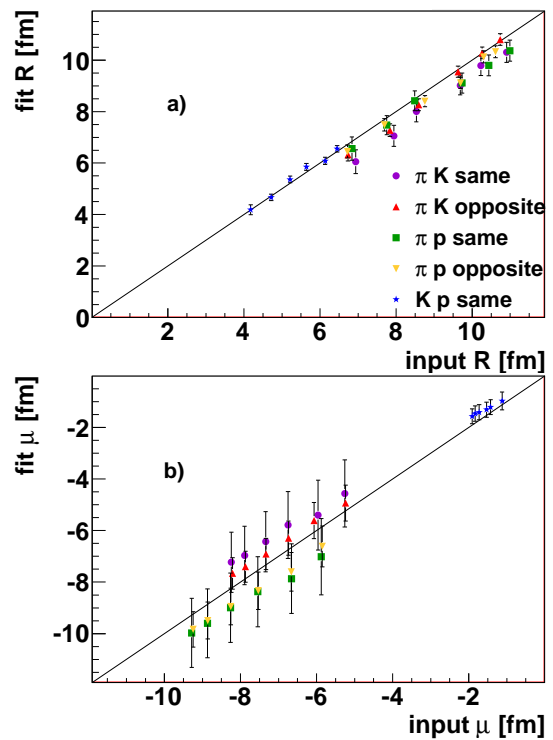}
\end{center}
\vspace{-6.5mm}
\caption{(Color on-line) Same as Fig.~\ref{fig:inpfitcor}, but the fit
procedure is done with adjusted purity values (see text for details).
\label{fig:inpfitcorpm}}
\end{figure}

One obvious candidate for the explanation of this shift is pair
purity $P$. The fit procedure assumes that {\it all} the pairs 
in the source come from the region with Gaussian density profile. On
the other hand, a realistic model, such as {\tt THERMINATOR}, clearly shows
that there are significant non-Gaussian long-range tails in the
separation distributions; this is illustrated in
Fig.~\ref{fig:longtail}. One can perform a simple calculation. Take
the model radius $R_{av}^{L}$ obtained from fitting the model
separation distribution with~\eqref{eq:gausjacob} and by integrating
this distribution one obtains the number of pairs $N_G$ within the
Gaussian core of the source. $N_G$ is then the area below the fit
curve on the lower--right panel of Fig.~\ref{fig:longtail}. This can be
compared to the total number of pairs $N_A$ obtained by simply
counting the number 
of model pairs, or in other words - the area below the points on the
lower--right panel of Fig.~\ref{fig:longtail}. The value $f_{FC}
= N_G/N_A$ is the number of ``femtoscopically correlated'' pairs and
should be simply treated as ``purity'' in the sense discussed in
Section~\ref{sec:purity}. One needs to correct for these
``non-gaussian'' effects. In order to do that one calculates $f_{FC}$
and then treats it as a fixed parameter in the fitting process. 
Note that the $f_{FC}$ estimation procedure by design takes into
account two effects, which cannot be easily disentangled - the pairs
in the long-range tails and the fact that the shape of the core system
is not a Gaussian in a 1D representation. We have 
performed the estimation, based on the input model pair separation
distributions and find that $f_{FC}$ depends both on the pair type and
centrality. The exact values are given in Tab.~\ref{tab:theorypur}.
These values are obviously model dependent and are strongly influenced
by the size of the long-range tails in the separation
distributions. It is well known that the source of such tails are, to
a large degree, particles coming from the strongly decaying
resonances. It is therefore important to use a model like {\tt
THERMINATOR}, which fully includes all known resonance propagation and
decay, to determine purity. A clear trend exists for all pair types:
the $f_{FC}$ get smaller as collisions get more peripheral. The
long-range resonance tails and non-Gaussian effects get relatively more
important as one moves away from central collisions.

The ``experimental fits'' have been redone, this time with ``purity''
fixed to the listed $f_{FC}$ values. The results are shown in
Fig.~\ref{fig:inpfitcorpm}. All results now lie close to the ``perfect''
$x=y$ line within the statistical error. This shows that, having in
mind the caveat mentioned above, the analysis technique is able to
produce reliable results, and that the results of the ``experimental
fitting'' procedure do indeed provide a valuable information about the
particle-emitting source. In particular, both the direction-averaged
size of the source and the emission asymmetry can be reliably
recovered. 

\begin{figure}[tb]
\begin{center}
\includegraphics[angle=0,width=0.48 \textwidth]{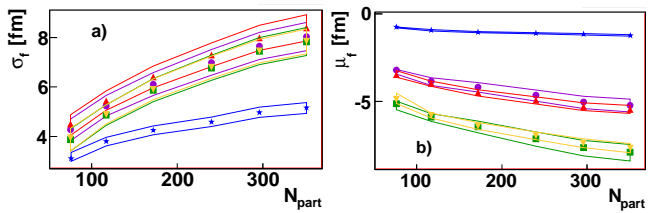}
\end{center}
\vspace{-6.5mm}
\caption{(Color on-line) Fitted parameters for all pair types,
versus centrality. System size ($\sigma$) in panel a), emission
asymmetry ($\mu$) in b). Colors and symbols are the same as in
Fig.~\ref{fig:inpfitcor}. The bands represent the change in the fit
parameters as the fixed purity $P$ is changed by $\pm 0.05$ for
pion--kaon and pion--proton pairs and $\pm 0.03$ for kaon--proton.
\label{fig:canpurerr}}
\end{figure}

The inherent uncertainty in the determination of $f_{FC}$ is the
source of the systematic error introduced by the method itself. In
Fig.~\ref{fig:canpurerr} we show how the fit value changes when purity
is varied within a reasonable range ($\pm 0.05$ for pion--kaon and
pion--proton and $\pm 0.03$ for kaon--proton). This results in the
following systematic uncertainties on the obtained values: for
same-sign pion--kaon $7\%$ on $\sigma$ and $7\%$ on $\mu$, for
opposite-sign pion--kaon $7\%$ on $\sigma$ and $4\%$ on $\mu$, for
same-sign pion--proton $8\%$ on $\sigma$ and $6\%$ on $\mu$, for
opposite-sign pion--proton $7\%$ on $\sigma$ and $4\%$ on $\mu$, for
same-sign kaon--proton $5\%$ on $\sigma$ and $4\%$ on $\mu$.

We note that the $f_{FC}$ estimation presented here is specific to
{\tt THERMINATOR} and represents the uncertainty within this model
itself. In addition, the absolute value as well as an estimated
uncertainty of $f_{FC}$ can be different in other models, e.g. in
rescattering codes. The total systematic uncertainty of the
experimental measurement should take this model dependence into
account. In addition, the experimental ``purity'', that is the
efficiency of particle identification, will also contribute to the
same uncertainty.  

\begin{figure}[tb]
\begin{center}
\includegraphics[angle=0,width=0.48 \textwidth]{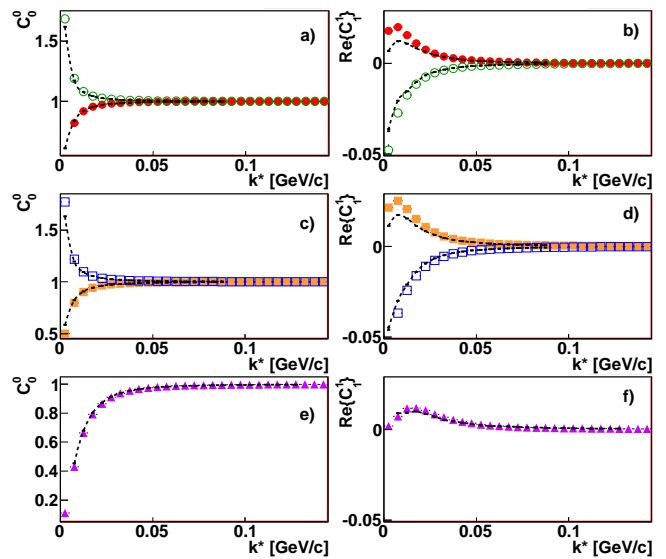}
\end{center}
\vspace{-6.5mm}
\caption{(Color on-line) Example of the correlation functions
calculated from {\tt THERMINATOR}+Lhyquid model for central Au+Au
collisions. Panels a), c), e) show the $C_0^0$ components, panels b),
d), f) show $\Re{C_1^1}$. Closed circles are same-sign pion--kaon,
open circles opposite-sign pion--kaon, closed squares are same-sign
pion--proton, open squares are opposite-sign pion--proton, triangles
are same-sign kaon--proton. Lines show the best-fit ``model''
correlation functions. 
\label{fig:canshfitcomp}}
\end{figure}

\section{Results of ``experimental-like'' analysis}
\label{sec:experimental}

\begin{figure}[tb]
\begin{center}
\includegraphics[angle=0,width=0.4 \textwidth]{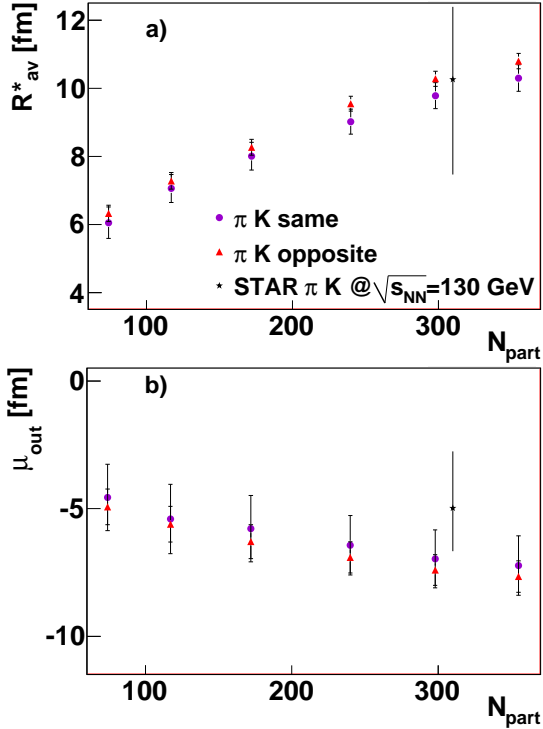}
\end{center}
\vspace{-6.5mm}
\caption{(Color on-line) The source parameters in PRF for pion--kaon pairs
obtained from the ``experimental'' fitting procedure. Panel a) shows
source size, b): emission asymmetry. Circles show same-sign pairs,
triangles: opposite sign. Open star is a STAR
measurement~\cite{Adams:2003qa} at $\sqrt{s_{NN}} = 130$~GeV, the
line represents statistical+systematic error (see text for details).  
\label{fig:pikdepfit}}
\end{figure}

In Section~\ref{sec:implement} we have described, in detail, the
experimental procedure to analyze the non-identical particle
correlations. For the model analysis of the {\tt THERMINATOR}
output, we have used the complete two-particle method for calculating
the correlation function and we have
obtained the pion--kaon, pion--proton and kaon--proton correlation
functions. They are, from a point of view of a formalism and our
analysis methods, identical to the correlation functions that one
might obtain in the experiment. We calculated sets of 1D
histograms which correspond to theses functions represented in
spherical harmonics. We neglected the strong interaction
component of the pair wave-function for simplicity and speed of
calculation. Then we switched this effect off in the
fitting procedure as well. Obviously, when fitting the true
experimental functions, one will use the full pair wave-function
calculation. The strong interaction effect is known to be small
compared to the Coulomb for the pairs of interest, so we do not expect
any systematic effect on the fit values coming from this
simplification. In addition, in the real experimental correlation
function, one expects some non-femtoscopic effects in addition to the
pure femtoscopic one. These need to be dealt
with on a case-by-case basis, as they will strongly depend on
experimental conditions. For this discussion we assume that any such
effects can be identified and that the experimental correlation
function can be properly corrected, so that only the femtoscopic
effect remains. 


\begin{figure}[tb]
\begin{center}
\includegraphics[angle=0,width=0.4 \textwidth]{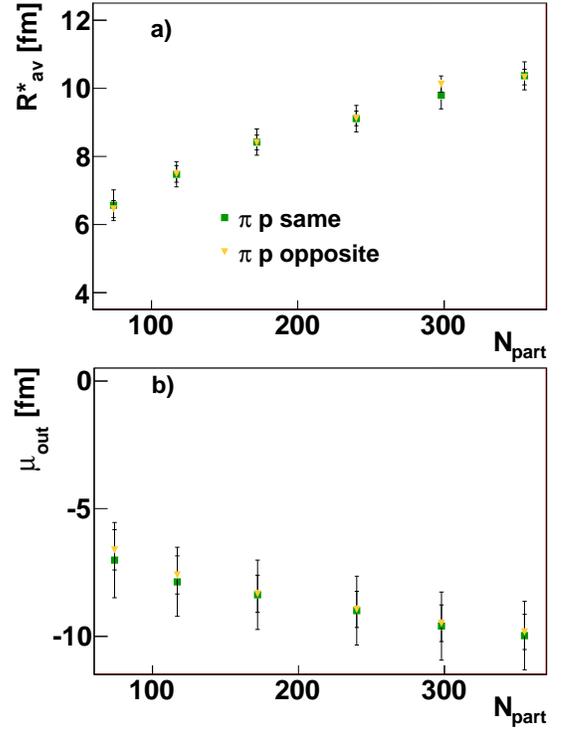}
\end{center}
\vspace{-6.5mm}
\caption{(Color on-line) The source parameters in PRF for pion--proton pairs
obtained from the ``experimental'' fitting procedure. Panel a): source
size, b): emission asymmetry. Squares show same-sign pairs, triangles:
opposite sign. 
\label{fig:pipdepfit}}
\end{figure}

We proceeded to treat the calculated correlation functions as if they
were coming from the experiment. No other information, except of the
pair purities is used in the procedure. For
purity correction we used values listed in Tab.~\ref{tab:theorypur}, as
we would have done in the real data analysis. The 
goal of the exercise it to confirm that the obtained ``experimental
fit values'' actually correspond to the true values, shown in the
previous Section and obtained directly from the emission functions.
The example correlation functions together with the fitted ``model''
ones are shown in Fig.~\ref{fig:canshfitcomp}. For pion--kaon and
pion--proton one can see the positive correlation effect for
opposite-sign pairs and a negative one for same-sign pairs. The
$\Re{C_1^1}$ components also show mirror effect. For kaon--proton the
correlation effects starts at larger $k^{*}$ (is wider) and is more
pronounced, as expected from a smaller Bohr radius for this pair. The
lines show the functions fitted with the {\tt CorrFit} program.

\begin{figure}[tb]
\begin{center}
\includegraphics[angle=0,width=0.4 \textwidth]{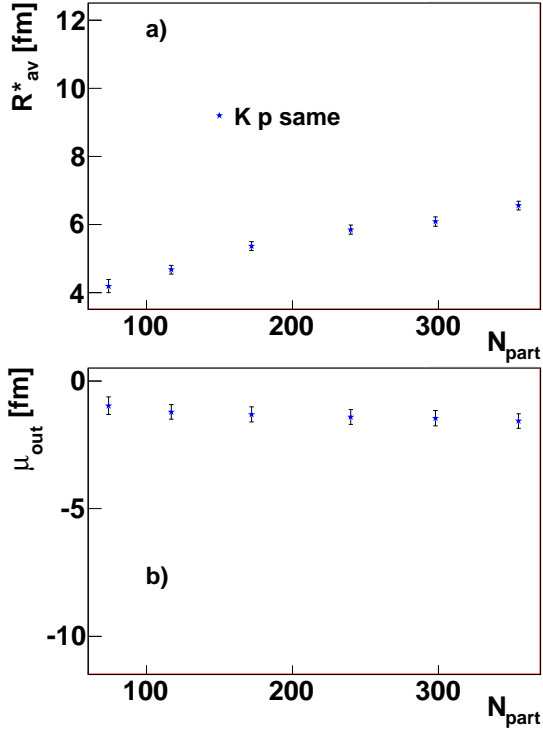}
\end{center}
\vspace{-6.5mm}
\caption{(Color on-line) The source parameters in PRF for kaon--proton pairs
obtained from the ``experimental'' fitting procedure. Panel a) shows
source size, b): emission asymmetry. 
\label{fig:kpdepfit}}
\end{figure}

The fitting described in Sec.~\ref{sec:nonidform} assumes that the
source is a 3D Gaussian in LCMS. The direct output of the fitting
procedure is a size of the system in the $out$ direction $\sigma^f$
and the emission asymmetry $\mu^f$ in that direction. From these values
one can calculate, via the relations specified in
Appendix~\ref{sec:emfuncs}, all the other source size
characteristics. In the fitting procedure each pair is treated
individually, so its velocity is known. One can therefore directly
determine the direction averaged source size in PRF - there is no need
to use averaged pair velocity in transformation from LCMS to PRF, like
in Eq.~\eqref{eq:lcmstoprfR}. In
Figs.~\ref{fig:pikdepfit},~\ref{fig:pipdepfit},~\ref{fig:kpdepfit} we
plot this directionally averaged radii in PRF $R^{*}_{av}$. Out of
the three equivalent values: $\sigma_f$, $R^{L}_{av}$ and $R^{*}_{av}$
one must choose one for presentation of the results. When we present
the results of our calculations, we chose $R^{*}_{av}$, as it is more
natural than $\sigma_f$ (which requires the knowledge of other
directions multipliers in order to be meaningful) and describes the
emission function in PRF, so there is no need for assumptions required
in the LCMS to PRF transformation. On the other hand, the 3D identical
particle femtoscopy analysis produces source sizes directly in LCMS,
so if we want to compare to these results, we will use
$R_{av}^{L}$. 

In Fig.~\ref{fig:pikdepfit} we show the ``experimental-like'' fit
results $R^{*}_{av}$ for pion--kaon pairs. This should be compared to
the ``true'' values from Fig.~\ref{fig:pikdep}. The correlation between
``true'' and ``fitted'' values is also shown directly in 
Fig.~\ref{fig:inpfitcorpm}. We can see that even with all the
assumptions and simplifications that are used in the fitting procedure
one is able to recover the true system size. The accuracy for
same-sign pion--kaon function is 8\%, for opposite-sign is 5\%, with
the largest deviation of 10\%. These are comparable to the statistical
error of the fit. For the asymmetry the ``input'' and ``fit'' values
were identical within 10\%. All results were in agreement within
statistical error of the fit. 

A single measurement is available for the pion--kaon system at RHIC,
done by STAR at $\sqrt{s_{NN}} = 130$~GeV~\cite{Adams:2003qa}. To
compare it with the values presented in Fig.~\ref{fig:pikdepfit} we
needed to account for two effects. The measurement was corrected for a
fraction of {\it non-primary} particles (particles not coming from
the primary vertex), which is an experimental correction. But, it was
not corrected for {\it non-correlated primary pairs}, a correction
which we described in this work. Taking the dependency in 
Fig.~\ref{fig:purdep} as a guideline and the purity estimate of 0.85
for pion--kaon in central Au+Au we have scaled the reported system size
and asymmetry accordingly. Moreover, the fitting performed
in~\cite{Adams:2003qa} assumed a specific shape (size equal in $out$,
$side$ and $long$ directions in PRF), from which we  recalculated
$R_{av}^{*}$, to be compared with results in this work. The
measurement was done at a 
colliding energy different than the one considered in this work, so we
plot the result at a corresponding $N_{part}$. One can see that the
system size is in perfect agreement with the model predictions in this
work; the asymmetry seems to be slightly smaller than the prediction,
but in agreement within statistical+systematic error.


In Fig.~\ref{fig:pipdepfit} the same results are shown for pion--proton
pairs. The accuracy of the system size determination is 5\%,
comparable to the statistical fit error. The asymmetry is reproduced
with worse quality, with discrepancies of up to 12\%, comparable to the
statistical error. 

In Fig.~\ref{fig:kpdepfit} the  results are shown for kaon--proton
pairs. The system size determination is good: 3\%, better
than the statistical fit error. The asymmetry is reproduced with
discrepancies of up to 15\%; however, since the absolute values are
small (compared to the pion--kaon and pion--proton case), the absolute
value of the difference is comparable with the systematic error of the
fit. We conclude that the system size and the emission asymmetry can
be reliably recovered, with the systematic error due to the procedure
itself less than 10\%. However, one must correctly determine the
systematic error due to the pair purity estimation as well. We note
that the tests of the method have been performed for Therminator model
only, so they are not necessarily general. However the tests were
done for many centralities (system sizes) and pair types and the
method was found to be working in all cases, so we have some
confidence that it should work for other models as well. One
possibility for further studies would be to perform similar tests with
a model that introduces particle rescatterings.


\begin{figure}[tb]
\begin{center}
\includegraphics[angle=0,width=0.4 \textwidth]{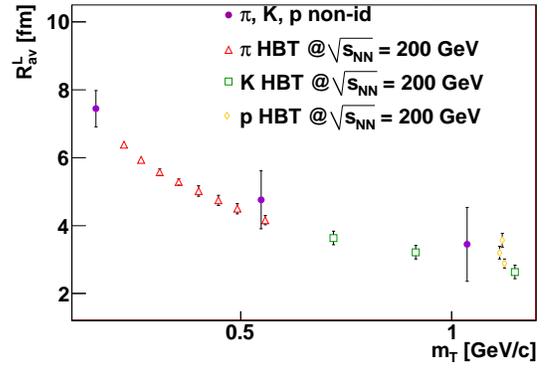}
\end{center}
\vspace{-6.5mm}
\caption{(Color on-line) Comparison of overall size obtained from the
non-identical particle calculation with the RHIC data for central Au+Au
collisions at $200$~GeV/c. Closed circles are (from left to right)
single particle pion, kaon and proton radii inferred from non-id
simulations (centrality 5-10\%); open triangles are pion radii from
STAR~\cite{collaboration-2005-71} (centrality 0-5\%), open squares are
pion radii from PHENIX~\cite{Afanasiev:2009ii} (centrality 0-30\%),
open diamonds are proton results from STAR~\cite{Gos:2007cj}
(centrality 0-10\%). 
\label{fig:mtdep}}
\end{figure}

One would also like to compare the results of non-identical particle
correlations with the wealth of data coming from femtoscopic analysis
at RHIC. This presents a complication, since the identical-particle
femtoscopy is usually presented in 3D form as single particle ``HBT
radii'' $R_{out}$, $R_{side}$ and $R_{long}$, while for analysis in
this work we only have averaged two-particle source size $\sigma^{f}$
and emission asymmetry $\mu^{f}_{out}$. From its definition we
conclude that e.g.\ $\sigma^{f}_{\pi K}$ should correspond to the
variance of convolution of two Gaussians: one with variance of
$R_{out}$ for pions and the other with variance of $R_{out}$ for kaons
at the same velocity. Both the identical particle $R$'s and the
non-identical $\sigma^{f}$ are defined in LCMS. Again we refer the
reader to Appendix~\ref{sec:emfuncs} for explicit relations between
the two. 
%
%
%
From the input data (either $R_{out}$, $R_{side}$ and $R_{long}$ in
LCMS for identical particles or $\sigma_f$ and $\mu_f$ for
non-identical) we calculate the directionally averaged source size
$R_{av}^{L}$. By comparing these values, defined in LCMS, as 
opposed to the $R_{inv}$ defined in PRF, we avoid the
unnecessary approximation coming from the determination of the
averaged pair velocity, needed for the determination of $\left <
\gamma \right >$. Such comparison is made in Fig.~\ref{fig:mtdep},
where data from most central Au+Au collisions are shown. The open
points are experimental data for pions, kaons and protons. The closed
points are single-particle source sizes inferred from fits to the
non-identical particle correlation functions calculated for {\tt
THERMINATOR}. One can see a very good agreement between the model
prediction and the data. All datapoints follow the ``$m_{T}$ scaling''
trend predicted by hydrodynamics. A direct comparison of
non-idnentical particle sizes and asymmetries between model and data
would be an even better test. However we stress that such comparisons
can only be made, provided that other observables related to source
dynamics, such as e.g. particle spectra, elliptic flow and identical
particle femtoscopy are reproduced in the model. These tests have been
performed for Therminator in earlier
works~\cite{Kisiel:2008ws,Broniowski:2008vp}. 

\begin{figure}[tb]
\begin{center}
\includegraphics[angle=0,width=0.4 \textwidth]{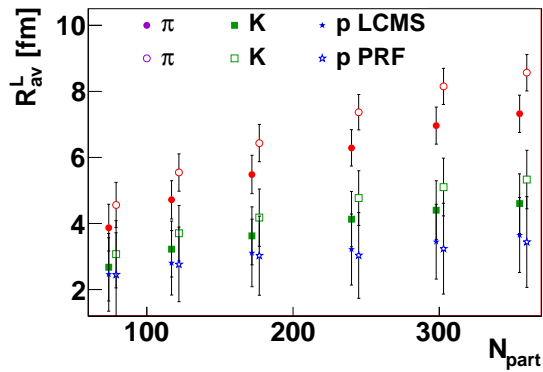}
\end{center}
\vspace{-6.5mm}
\caption{(Color on-line) The single-particle source sizes inferred from
the non-identical simulations versus centrality. Closed points are for
LCMS sizes, open (shifted for clarity) for PRF. Circles are pions,
squares: kaons, stars: protons.  
\label{fig:singler}}
\end{figure}

We have calculated the single-particle source sizes for all
centralities and plotted them in Fig.~\ref{fig:singler}. An ordering
of source sizes with particle mass is seen as expected, as well as
increase of the overall source size with $N_{part}$. Sizes for PRF are
larger, as the $out$ radius is scaled by the $\left < \gamma_t \right
>$ factor. 

\begin{figure}[tb]
\begin{center}
\includegraphics[angle=0,width=0.4 \textwidth]{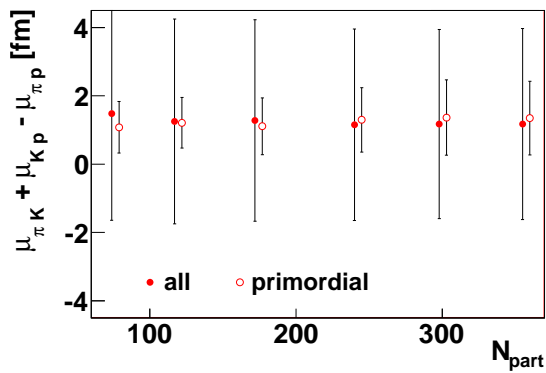}
\end{center}
\vspace{-6.5mm}
\caption{(Color on-line) Emission asymmetry consistency cross-check 
versus centrality. Closed points are for all particles, open (shifted
for clarity) for primordial only. 
\label{fig:msum}}
\end{figure}

As a final cross-check we show in Fig.~\ref{fig:msum} that the
emission asymmetries do add up to zero within the systematic error of
the measurement. 

\section{Conclusions}
\label{sec:conclusions}

The {\tt THERMINATOR}+Lhyquid model was briefly introduced, and shown to
have a unique set of features: hydrodynamic expansion and inclusion of
resonance propagation and decay for all known particle types. Both
features are important for non-identical particle
femtoscopy. Simulations with this model provided predictions for
emission asymmetries between particles of different masses, and
validated the analysis methods of non-identical particle femtoscopy.

We underlined the importance of the emission asymmetry measurement and
provided predictions for RHIC energies. We showed that ``emission
asymmetry'', or the non-zero difference in the mean emission points of
particles of same velocity but different mass is predicted to arise in
hydrodynamical calculation. It was shown to be intimately related to
the collective behavior of matter and to be a direct consequence of the
$x-p$ correlations. It is the most direct and unambiguous femtoscopic
signal of such behavior, and provides an independent strong
constraint on the models aiming to describe the space--time evolution
of the heavy-ion collision. Specific predictions, both qualitative and
quantitative, have been given for such asymmetries, for all the
considered pair types (pion--kaon, pion--proton, kaon--proton). It was
shown that the lighter particle is always emitted closer to the center
of the source, giving a negative emission asymmetry $\mu_{out}$ in the
$out$ direction. Predictions for the size of the source, including the
dependence on centrality, were also given, which were cross-checked
with the more precise results from identical particle femtoscopy. 

The influence of strongly decaying resonances on emission asymmetries
was also studied in detail. It was shown that due to a specific
combination of decay kinematics' properties for particles of different
masses, the emission asymmetry produced by an earlier hydrodynamic
stage is further magnified. At the same time the resonance decay
process alone produces emission asymmetry an order of magnitude smaller
than the $x-p$ correlation, so it cannot be used as an alternative
explanation of the phenomenon.

We have presented, in some detail, the theoretical basis and some
technical aspects of the non-identical particle femtoscopy, to be used
in ultra-relativistic heavy-ion collisions. In particular it was shown
how to access the ``emission asymmetry'', a piece of femtoscopic
information that can only be accessed via this type of analysis. It
was pointed out that spherical harmonics representation of the
correlation function has useful synergies with the analysis; the use
of such representation was strongly advocated. Important
corrections to the correlation function were identified: pair purity
and momentum resolution, and experimentalist recipes were given for
applying them to the real data. Extracting femtoscopic information from
the non-identical particle correlation function requires a numerical 
fitting procedure. It was described in detail, and implemented as a
computer code. It was also shown that, after application of the proper
purity correction, the fit procedure was able to recover the ``true''
model input values, which was an important validation of the method.

An estimate of the ``fraction of correlated pairs'', coming from
significant non-Gaussian shape and long-range tails in the pair
separation distributions, has been obtained, based on the {\tt
THERMINATOR} simulation. The value can be an input to an experimental
analysis of RHIC data, however one must keep in mind that it is model
dependent. The systematic error coming from this theoretical estimate
was also given.

Finally, the ``experimentalists'' analysis procedure, described in the
paper, was applied to the correlation functions obtained from the
model calculations. It was shown that the method reliably recovers the
input model values in realistic conditions. Internal consistency
cross-checks were proposed and tested. The way to compare results of
non-identical particle femtoscopy and the femtoscopic sizes from
identical particle analysis was presented; the {\tt THERMINATOR}
predictions were shown to be in agreement with available identical
particle HBT radii results from RHIC. 

\section*{Acknowledgements}

I would like to thank Richard Lednicky, Wojciech Florkowski and Yiota
Foka for comments and very helpful discussions. 

\appendix

\section{Emission function measures}
\label{sec:emfuncs}

In femtoscopy one considers pairs of particles. Of particular interest
are observables related to their relative separation $\vec r$. We
refer the reader to the beginning of Section~\ref{sec:modelpreditions}
for the description of symbol conventions, the reference system and
the relevant reference frames. The transition from LCMS to PRF is 
simply the boost along the outwards direction, with the transverse
velocity of the pair $\beta_t$:
\begin{eqnarray}
r^{*}_{out} &=& \gamma_t \left ( r_{out} - \beta_t \Delta t \right )
\nonumber \\
r^{*}_{side} &=& r_{side} \nonumber \\
r^{*}_{long} &=& r_{long} \nonumber \\
\Delta t^{*} &=& \gamma_t \left ( \Delta t - \beta_t r_{out} \right ) .
\label{eq:rtorstar}
\end{eqnarray}
However, in our calculation we always use the equal time approximation,
which means that we neglect the time difference $\Delta t^{*}$ in the
PRF. From the components we also calculate the length of the relative
separation vector:
\begin{equation}
r^{*} \equiv r_{inv} = \sqrt{{r^{*}_{out}}^2 + {r^{*}_{side}}^2 +
{r^{*}_{long}}^2} .
\label{rinvdef}
\end{equation}

The two-particle emission function is, from a mathematical point of
view, any function of the separation four-vector, $S(\Delta x)$, where
$\Delta x = \{\Delta t,r \}$ can be expressed in any reference frame. In
femtoscopy we use 
specific functional forms of $S$, which are characterized by
parameters. One must keep in mind that even though the relation between
{\it relative separation $\vec r$} values in LCMS and PRF is given by
simple equation Eq.~\eqref{eq:rtorstar}, the relations between {\it
source function parameters} ($R$'s and $\mu$) in these reference
frames may not be so simple. Moreover, even if there exists a simple
parametrization of the source in 3D variables ($r_{out}$, $r_{side}$,
$r_{long}$), it does not mean that this source distribution expressed
in the magnitude of the relative separation $r$ will also have a simple
parametric form. The aim of this Appendix is to derive relations
between commonly used source function parameters, defined in various
reference frames as well as between 3D and 1D representation parameters. 
Similar considerations, for different source parametrizations have
been done in the past~\cite{Chacon:1991ri}.

We must also repeat the description of Eq.~\eqref{eq:sfromconv} where we
state that the two-particle emission function $S_{AB}$ is (in case of
independent emission) a convolution of two single particle emission
functions. Let us emphasize this complication of the femtoscopic
measurement: we are interested in the {\it single particle} emission
functions, as they directly characterize the source. On the other hand
femtoscopy provides information about the {\it two-particle} (or
{\it relative)} emission functions only. Moreover, the convolution procedure
of Eq.\eqref{eq:sfromconv} is not reversible. It means that, in a
strict mathematical sense, one cannot recover the full information
about the individual single particle emission functions by measuring
only the separation distribution. Nevertheless, in femtoscopic
measurements, reasonable additional assumptions can be made and
we are able to recover some of the information. In this Appendix we
will discuss how to extract the single-particle emission function
parameters from the measured two-particle ones.

\subsection{Traditional emission function parametrizations}

We list here the traditional functional forms of the emission function
that are used in femtoscopy and list their parameters. The easiest is
the 1-dimensional ``$R_{inv}$'' parametrization:
\begin{eqnarray}
S_{1d}^{P}(\vec r^{*}) &\equiv& \frac{dN} {d^3 \vec r} = \exp \left 
(-\frac{{r^{*}_{out}}^2+{r^{*}_{side}}^2+{r^{*}_{long}}^2} {4 
{R_{inv}}^2}  \right ) \nonumber \\
S_{1d}^{P}(r^{*}) &\equiv& \frac {dN} {dr^{*}} = {r^{*}}^2 \exp \left ( -
\frac {{r^{*}}^2} {4 {R_{inv}}^2} 
\right ) .
\label{eq:srinv}
\end{eqnarray}
Note that the formulas neglect the $\Delta t^{*}$ dependence, which 
is possible thanks to the equal-time approximation in PRF, allowing
one to neglect $\Delta t^{*}$ dependence of $\left | \Psi \right |^2$.
The change from the 3D variables to the 1D one requires the introduction
of the proper Jacobian. Let us also explain the factor of 4 before the
``Gaussian sigma'' $R_{inv}$ instead of the usual 2. It is a
particular property of a Gaussian distribution that a convolution of
two Gaussians is also a Gaussian with its sigma being the quadratic
sum of the sigmas of the individual distributions. The $R_{inv}$
parametrization is used for identical particles, for which
Eq.~\eqref{eq:sfromconv} can be simplified by noticing that both
emission functions are the same. If ones assumes that a
single-particle emission function is a Gaussian, then the two-particle
one is also a Gaussian with $\sigma$ multiplied by
$\sqrt{2}$. Therefore, by fitting the {\it two-particle} distribution 
with the functional form~\eqref{eq:srinv} one conveniently obtains the
{\it single-particle} $\sigma$: the $R_{inv}$. Therefore, the $R_{inv}$
``HBT radii'' reported by experiments are the single-particle
emission function Gaussian widths, where the emission function is
assumed to be a 3D Gaussian in the PRF.

A more sophisticated emission function form, used by all RHIC and SPS
experiments to report identical pion femtoscopy results is: 
\begin{equation}
S_{3d}^{L}(\vec r) = \exp \left (-\frac{{r_{out}}^2} {4 {R^{L}_{out}}^2}
-\frac{{r_{side}}^2} {4 {R^{L}_{side}}^2} -\frac{{r_{long}}^2} {4
{R^{L}_{long}}^2}  \right ) .
\label{eq:soutsidelong}
\end{equation}
We again note the factor of 4 before the $\sigma$'s, making them the
single-particle widths. The important difference is that this emission
function has three independent width $R^{L}_{out}$, $R^{L}_{side}$ and
$R^{L}_{long}$ and they are defined in the LCMS (hence the $L$
superscript), not in PRF. Note that, unlike in PRF, we do not use
equal-time approximation in LCMS. For identical particles this is not
a problem: only the Coulomb part of the wave-function depends on
$\Delta t$, but it is factorized out in the usual fitting
procedures, and replaced by the averaged value. For non-identical
particles one can adopt one of two approaches. One can continue to use
Eq.~\eqref{eq:soutsidelong} to be consistent with identical particle
femtoscopy, but in this case the $\Delta t$ spread will get absorbed into the
three spatial radii. Or one can add the fourth component with the time
spread to Eq.~\eqref{eq:soutsidelong}. Since we aim to compare the
results from non-identical and identical particle analysis and since
we are in any case only able to recover the dimensionally-averaged
source size for non-identical particles we chose the first solution.

For non-identical particles we must make one additional modification,
the mean emission point should be allowed to differ from zero, at
least in the $out$ direction. Moreover the simple connection between
the single- and two-particle sizes is no longer possible, since the
underlying single-particle emission functions are now different:
\begin{equation}
S_{3d}^{L,N}(\vec r) = \exp \left (-\frac{(r_{out} - \mu^{L}_{out})^2} {2
{R^{L}_{out}}^2} -\frac{{r_{side}}^2} {2 {R^{L}_{side}}^2}
-\frac{{r_{long}}^2} {2 {R^{L}_{long}}^2}  \right ) .
\label{eq:soutsidelong}
\end{equation}
We introduced an additional parameter: the mean of the distribution in
the $out$ direction $\mu^{L}_{out}$ and the source sizes are now the
{\it two-particle} ones. 

In order to facilitate the comparison between 1D and 3D source sizes
in LCMS we also introduce the emission function with one,
directionally averaged, source size in LCMS $R_{av}^{L}$:
\begin{equation}
S_{1d}^{L,N}(\vec r) = \exp \left (-\frac{(r_{out} - \mu^{L}_{out})^2 +
{r_{side}}^2 + {r_{long}}^2} {2 {R^{L}_{av}}^2} \right ) .
\label{eq:savl}
\end{equation}

\subsection{Relating 1D and 3D source sizes}

The non-identical particle femtoscopy has been, so far, limited to the
measurement of the directionally averaged source size $R^{L}_{av}$. A
question arises, what is the relation between $R^{L}_{av}$ and
three-dimensional source sizes $R_{out}$, $R_{side}$ and
$R_{long}$. Let us write explicitly the form of the emission function
in magnitude of $r$: 

\begin{eqnarray}
S^{L}_{3d}(r) &=& \int \exp \left( -\frac {r_{out}^2} {2 {R^{L}_{out}}^2}
-\frac {r_{side}^2} {2 {R^{L}_{side}}^2} -\frac {r_{long}^2} {2
{R^{L}_{long}}^2} \right) \nonumber \\ 
& & \times \delta(r- \sqrt{r_{out}^2 + r_{side}^2 + r_{long}^2})
dr_{out} dr_{side} dr_{long} .\nonumber \\
\label{eq:sfromosl}
\end{eqnarray}
To find the 1D source size corresponding to the 3 3D ones, we assume
that the above distribution can be approximated by:
\begin{equation}
S^{L}_{1d}(r) = r^2 \exp \left ( -\frac {r^2} {2 {R_{av}^{L}}^2} \right).
\label{eq:sfromav}
\end{equation}
First, let us note that $S^{L}_{3d}$ simplifies exactly to $S^{L}_{1d}$ only in
the special case of $R^{L}_{out}=R^{L}_{side}=R^{L}_{long}$, in which
case $R^{L}_{av}=R^{L}_{out}$. If this condition is not met, the 
$S^{L}_{3d}$ is {\it not} a Gaussian in $r$, and the exact mathematical
relation between the 1D and 3D sizes does not exist. Nevertheless, for
realistic values of radii, $S^{L}_{3d}$ is not very different from a 
Gaussian and can be well approximated by $S^{L}_{1d}$. One can find an
effective approximate relation between $R^{L}_{av}$ and ($R^{L}_{out}$,
$R^{L}_{side}$, $R^{L}_{long}$) numerically in the following way. One
generates a significant sample of triplets ($r_{out}$, $r_{side}$,
$r_{long}$) where $r_{out}$ is randomly generated from a Gaussian of
width $R^{L}_{out}$, $r_{side}$ with width $R^{L}_{side}$ etc. Then,
one constructs a distribution $S(\sqrt{r_{out}^2 + r_{side}^2 +
r_{long}^2})$, to which one then fits numerically the functional
form~\eqref{eq:sfromav}. The results of the fit, $R^{L}_{av}$ is the
approximate 1D source size that we seek. We may also want to restrict
the fit range to small values of $r$ in order to minimize the
dependence on non-Gaussian features which will mostly affect the
large-$r$ region. We have performed such calculations and concluded
that the approximate relation is: 
\begin{equation}
R^{L}_{av} = \sqrt{({R^{L}_{out}}^2 + {R^{L}_{side}}^2 +
{R^{L}_{long}}^2)/3} .
\label{eq:ravfromosl}
\end{equation}
Note that the pair velocity does not enter into the derivation, so it is
equally valid for LCMS and PRF.

In addition, in non-identical particle femtoscopy one is able to
access the first moments of the source distribution. Then, $S$  
has the general form of: $S(R^{L}_{out}, R^{L}_{side}, R^{L}_{long}, 
\mu^{L}_{out}, \mu^{L}_{side}, \mu^{L}_{long})$. Due to symmetry
relations $\mu^{L}_{side}$ must vanish. For collider experiments with
symmetric rapidity acceptance $\mu^{L}_{long}$ vanishes as well, which
leaves one additional parameter $\mu^{L}_{out}$. By performing a
numerical procedure very similar to the one in the previous paragraph
(the only difference being that the $r_{out}$ is now randomly
generated from a Gaussian with the mean of $\mu^{L}_{out}$) we
obtained an equivalent approximate effective relation: 
\begin{equation}
R^{L}_{av} = \sqrt{({R^{L}_{out}}^2 + {R^{L}_{side}}^2 +
{R^{L}_{long}}^2)/3 + 0.3{\mu^{L}_{out}}^2} .
\label{eq:ravfromoslm}	
\end{equation}

\subsection{Relating single-particle and two-particle sizes}

As already mentioned for identical particles the relation between
two-particle and single particle sizes is a trivial $\sqrt{2}$ factor
if the assumption of a Gaussian emission function is made. For
non-identical particle sizes such simple connection is not possible,
and even an approximate one can only be made after certain
simplifications are done. Let us assume that the two particle types,
$A$ and $B$ are emitted according to the emission functions $S_{A}$
and $S_{B}$ which are Gaussians. The two particle emission function is
then:
\begin{eqnarray}  
S_{AB} (\vec r) &=& \int \exp \left ( -\frac {(x^{A}_{out} -
\mu^{A}_{out})^2} {2 {R^{A}_{out}}^2} -\frac {{x^{A}_{side}}^2} {2 
{R^{A}_{side}}^2} -\frac {{x^{A}_{long}}^2} {2 {R^{A}_{long}}^2}
\right ) \nonumber \\
& \times & \exp \left ( -\frac {(x^{B}_{out} -
\mu^{B}_{out})^2} {2 {R^{B}_{out}}^2} -\frac {{x^{B}_{side}}^2} {2 
{R^{B}_{side}}^2} -\frac {{x^{B}_{long}}^2} {2 {R^{B}_{long}}^2}
\right ) \nonumber \\
& \times & \delta (r_{out} - x^{A}_{out} + x^B_{out}) dx^A_{out}
dx^B_{out} \nonumber \\
& \times & \delta (r_{side} - x^{A}_{side} + x^B_{side}) dx^A_{side}
dx^B_{side} \nonumber \\ 
& \times & \delta (r_{long} - x^{A}_{long} + x^B_{long}) dx^A_{long} 
dx^B_{long} .
\label{eq:sabfromsingle}
\end{eqnarray}
Performing the integration and neglecting the unimportant
normalization constants one obtains:
\begin{eqnarray}
S_{AB} &=& \exp \left ( - \frac {[r_{out} - (\mu^{A}_{out} - 
\mu^{B}_{out})]^2} {2 ({R^{A}_{out}}^2 + {R^{B}_{out}}^2) } \right )
\nonumber \\ 
&\times& \exp \left ( - \frac {r_{side}^2} {2 ({R^{A}_{side}}^2 +
{R^{B}_{side}}^2) } \right ) \nonumber \\ 
&\times& \exp \left ( \frac {r_{long}^2} {2 ({R^{A}_{long}}^2 +
{R^{B}_{long}}^2) }  \right ) 
\label{eq:deftwo}
\end{eqnarray}
giving us immediately $\mu^{AB}_{out} = \mu^{A}_{out} - \mu^{B}_{out}$
and $R^{AB}_{x} = \sqrt{{R^{A}_{x}}^2 +{R^{B}_{x}}^2}$, as
expected. Obviously, one cannot recover the two single-particle source
sizes from the one two-particle size. However, in this work we have
calculated three independent two-particle sizes: for pion--kaon,
pion--proton and kaon--proton systems. Therefore we have a set of three
equations~\eqref{eq:deftwo}, with $A$ and $B$ being $\pi$ and $K$ in
the first one, $\pi$ and $p$ in the second one and $K$ and $p$ in the
third one respectively. This set of equations has three unknowns, the
single particle source sizes: $R^{\pi}$, $R^{K}$ and $R^{p}$, so we
can solve it to calculate them. 

The procedure has to be carried out in a few steps. We
start with the fit values $\sigma_f$. From this we calculate the
approximate overall averaged source size $R^{av}_f$, following our fit
assumptions: $R_{out} = \sigma_f$, $R_{side}= \sigma_f$ and $R_{long}
= 1.3 \sigma_f$. With these values we write the set of equations for
the particle source sizes:
\begin{eqnarray}
\sigma^{\pi K}_f &=& \sqrt{{\sigma^{\pi}_f}^{2} + {\sigma^K_f}^{2}}
\nonumber \\ 
\sigma^{\pi p}_f &=& \sqrt{{\sigma^{\pi}_f}^{2} + {\sigma^p_f}^{2}}
\nonumber \\ 
\sigma^{K p}_f &=& \sqrt{{\sigma^{K}_f}^{2} + {\sigma^p_f}{2}} 
\label{eq:twosigmas}
\end{eqnarray}
and solve it obtaining:
\begin{eqnarray}
\sigma^{\pi}_f &=& \sqrt{( {\sigma^{\pi K}_f}^2 + {\sigma^{\pi p}_f}^2
- {\sigma^{K p}_f}^2)/2} \nonumber \\
\sigma^{K}_f &=& \sqrt{( {\sigma^{\pi K}_f}^2 - {\sigma^{\pi p}_f}^2
+ {\sigma^{K p}_f}^2)/2} \nonumber \\
\sigma^{p}_f &=& \sqrt{( {-\sigma^{\pi K}_f}^2 + {\sigma^{\pi p}_f}^2
+ {\sigma^{K p}_f}^2)/2} .
\label{eq:singlefromtwo}
\end{eqnarray}
From these fit parameters one then calculates directionally averaged
single-particle radius $R_{av}^{L}$ with Eq.~\eqref{eq:ravfromosl} -
$\sigma_f$ is treated as $R^{L}_{out}$. These can be compared to the
$R^{L}_{av}$ directionally averaged radii in LCMS, which can be
calculated with Eq.~\eqref{eq:ravfromosl} from the {\it
single-particle} 3D HBT radii published by the RHIC experiments.

\subsection{Dealing with non-Gaussian source functions}

Let us now consider a non-Gaussian source, as it is evident,
e.g.\ from this work, that realistic models predict that the emitting 
source will not have a perfect Gaussian shape. However, the 
two-particle emission function, defined as a convolution of two
single-particle sources, is required to be closer to Gaussian than the
single-particle functions. And the single particle functions
themselves, even though non-Gaussian, are not very different: they
usually have a large peak and some long-range tails. From this we
conclude that even though we know that the two-particle source
function is not Gaussian, we may assume that it will have
Gaussian-like features: it will have a peak, and the distribution
around that peak can be reasonably approximated by a Gaussian. It will
also usually have long-range non-Gaussian tails. This means that we
can define two variables that can be used as measures of the variance:
either the RMS of the distribution, or the $\sigma$ of a Gaussian fit
to the peak structure in the emission function. For the study of
non-identical particle averaged sizes $\sigma$ is the proper variable
to be used. That is because pair wave functions, which are the 
source of femtoscopic correlation, contain strong structures
near zero $r^{*}$, while at large $r^{*}$ they do not produce any
correlation. On the other hand the RMS of the distribution will be
very sensitive to the long-range tails, which is
undesirable in our case. Therefore we define that a ``femtoscopic
size'' $R$ for a non-Gaussian emission function is simply the $\sigma$
of a Gaussian fitted to the two-particle emission function near its
peak.  

\subsection{Obtaining $R_{av}$ from models}

In models we know the emission function exactly. Therefore we can
infer $R_{av}$ directly from it. We do this numerically, in a
manner very similar to the one shown in previous sections. First we
determine the probability distributions for various components of $r$:
$f_{o}(r_{out})$, $f_{s}(r_{side})$, $f_{l}(r_{long})$. For the
{\tt THERMINATOR} calculation, discussed in this work, we have found that
$f_{o}$ has the form of~\eqref{eq:fehas}, $f_s$ is Gaussian, and $f_l$
has the form~\eqref{eq:feh}. Now we proceed with the generation of
triplets, construction of the $S(r)$ and numerical fitting to obtain
$R_{av}$. Such $R_{av}$ can then be directly compared to the $R_{av}$
obtained from the ``experimental'' fitting procedure described above.

\subsection{Relations between source sizes in LCMS and PRF}

We can write approximate relations between source sizes in the two
reference systems, they are:

\begin{eqnarray}
R^{*}_{out} &=& R_{out} \left < \gamma_t \right > \nonumber \\
R^{*}_{side} &=& R_{side}  \nonumber \\
R^{*}_{long} &=& R_{long}  \nonumber \\
\mu_{out} &=& \left < \gamma_t \right > \mu^{L}_{out} .
\label{eq:lcmstoprfR}
\end{eqnarray}

Using Eqs.~\eqref{eq:ravfromosl},~\eqref{eq:ravfromoslm} one can obtain
the  relations for the averaged 1D radius $R^{*}_{av}$ as well. One
must also independently determine the $\left < \gamma_t \right >$ for
the pair sample used in a given analysis. 

\bibliography{ref-rr}

\end{document}